\documentclass[aps,pra,twocolumn,superscriptaddress,showkeys]{revtex4-1}
\usepackage[final]{graphicx}
\usepackage{epsfig}
\usepackage{amssymb}
\usepackage{amsmath}
\usepackage{bm}

\usepackage{cases}
\usepackage{multirow}
\usepackage{color}
\usepackage{ulem}
\usepackage{dcolumn}

\begin{document}

\title{Suppression of Black-body Radiation Induced Zeeman Shifts in the Optical Clocks \\
due to the Fine-structure Intramanifold Resonances}

\author{Zhi-Ming Tang}
\email{zmtang18@fudan.edu.cn}
\thanks{\\These authors contributed equally to this work.}
\affiliation{Shanghai EBIT Laboratory, Key Laboratory of Nuclear Physics and Ion-Beam Application (MOE), \\
Institute of Modern Physics, Fudan University, Shanghai 200433, China}

\author{Yuan-Fei Wei}
\thanks{These authors contributed equally to this work.}
\affiliation{State Key Laboratory of Magnetic Resonance and Atomic and Molecular Physics, Innovation Academy for \\
Precision Measurement Science and Technology, Chinese Academy of Sciences, Wuhan 430071, China}
\affiliation{Key Laboratory of Atom Frequency Standards, Innovation Academy for Precision Measurement \\
Science and Technology, Chinese Academy of Sciences, Wuhan 430071, China}
\affiliation{University of Chinese Academy of Sciences, Beijing 100049, China}

\author{B. K. Sahoo}
\affiliation{Atomic, Molecular and Optical Physics Division, Physical Research Laboratory, Navrangpura, Ahmedabad 380009, India}

\author{Cheng-Bin Li} \email{cbli@apm.ac.cn}
\affiliation{State Key Laboratory of Magnetic Resonance and Atomic and Molecular Physics, Innovation Academy for \\
Precision Measurement Science and Technology, Chinese Academy of Sciences, Wuhan 430071, China}

\author{\\Yang Yang} \email{yangyang@fudan.edu.cn}
\author{Yaming Zou}
\affiliation{Shanghai EBIT Laboratory, Key Laboratory of Nuclear Physics and Ion-Beam Application (MOE), \\
Institute of Modern Physics, Fudan University, Shanghai 200433, China}

\author{Xue-Ren Huang}
\affiliation{State Key Laboratory of Magnetic Resonance and Atomic and Molecular Physics, Innovation Academy for \\
Precision Measurement Science and Technology, Chinese Academy of Sciences, Wuhan 430071, China}
\affiliation{Key Laboratory of Atom Frequency Standards, Innovation Academy for Precision Measurement \\
Science and Technology, Chinese Academy of Sciences, Wuhan 430071, China}
\affiliation{Wuhan Institute of Quantum Technology, Wuhan 430206, China}

\date{21 September 2023}

\begin{abstract}
The roles of the fine-structure intramanifold resonances to the Zeeman shifts caused by the blackbody radiation (BBRz shifts) in the optical clock
transitions are analyzed. The clock frequency measurement in the $^1S_0-^3P_0$ clock transition of the singly charged aluminium ion (Al$^+$) has
already been reached the $10^{-19}$ level at which the BBRz effect can be significant in determining the uncertainty. In view of this, we probe
first the BBRz shift in this transition rigorously and demonstrate the importance of the contributions from the intramanifold resonances explicitly.
To carry out the analysis, we determine the dynamic magnetic dipole (M1) polarizabilities of the clock states over a wide range of angular
frequencies by employing two variants of relativistic many-body methods. This showed the BBRz shift is highly suppressed due to blue-detuning of the BBR spectrum to
the $^3P_0-^3P_1$ fine-structure intramanifold resonance in Al$^+$ and it fails to follow the usually assumed static M1 polarizability limit
in the estimation of the BBRz shift. The resonance also leads to a reversal behavior of the temperature dependence and a cancellation in the shift. After learning this
behavior, we extended our analyses to other optical clocks and found that these shifts are of the order of micro-hertz leading to fractional shifts
in the clock transitions at the $10^{-20}$ level or below.
\end{abstract}

\maketitle

\section{\label{sec:introduction}Introduction}

Cutting-edge optical clocks have achieved the fractional uncertainty and instability at $10^{-18}$ level or below \cite{Bloom-Nature-2014,
Huntemann-PRL-2016, Schioppo-nphoton-2017, Brewer-PRL-2019, Boulder-Nature-2021}. This progress makes the optical clocks the ideal candidates for
redefining the SI second \cite{Riehle-Metrologia-2018, Milner-PRL-2019} and enables them for many important applications, such as probing relativistic
geodesy \cite{McGrew-Nature-2018}, utilising for deep space search and global navigation \cite{Schuldt-GPSsolut-2021, Burt-Nature-2021},
searching for dark matters \cite{Kennedy-PRL-2020, Kobayashi-PRL-2022}, and testing fundamental physics \cite{Safronova-RMP-2018, Sanner-Nature-2019,
Lange-PRL-2021, Bothwell-Nature-2022}. The currently available high-performance optical clocks are generally based on either neutral atomic ensemble
in an optical lattice \cite{Derevianko-RMP-2011} or a singly charged ion in a radio-frequency trap \cite{Ludlow-RMP-2015}. Recently, another class of optical
clocks based on the highly charged ions (HCIs) \cite{Kozlov-RMP-2018, Yu-review-2023} have also been realized \cite{King-Nature-2022}. This has motivated
many groups to propose a wide range of HCIs as the prospective atomic clock candidates \cite{Derevianko-PRL-2012, Yudin-PRL-2014, Safronova-PRL-2014,
Yu-PRA-2016, Yu-PRA-2018, Yu-PRA-2019, Bekker-ncommun-2019, Lyu-arxiv-2023, Yu-arxiv-2023}. With the advent of modern laser technologies for quantum
manipulating of atoms and ions, improving accuracy of the optical clocks have been steadily gaining the ground. This continuously pushes the limits of
accuracy of frequency metrology and are anticipated to reach $10^{-21}$ precision in the coming two decades \cite{Derevianko-QST-2022}. In all the aforementioned clocks, it is also very important to estimate the subtle frequency shifts caused by the environment and apparatus to improve
their accuracy further.

Among the so far realized atomic clocks in the laboratory, the singly charged aluminum ion (Al$^+$) clock has been making continuous progress in the last
15 years \cite{Rosenband-Science-2008, Chou-PRL-2010, Brewer-PRL-2019}, thanks to the state-of-the-art sympathetic cooling and quantum-logic technologies.
The Al$^+$ clock developed by the NIST group reached the uncertainty in the frequency measurement within $10^{-18}$ level in 2010 \cite{Chou-PRL-2010}
and was improved to the $10^{-19}$ level in 2019 \cite{Brewer-PRL-2019}. The advantage of considering Al$^+$ as clock is primarily due to its
intrinsically ultra-narrow $3s^2~^1S_0 - 3s3p~^3P_0$ transition, which has a high quality factor of $1.45\times10^{17}$ \cite{Rosenband-PRL-2007}.
In addition, the clock transition has a small differential static electric dipole ($E1$) polarizability, $\Delta\alpha(0)=0.426(57)$ in atomic units (a.u.)
\cite{Brewer-PRL-2019, Wei-CPB-2022}, which causes a very low blackbody radiation (BBR) shift at the room temperature \cite{Kallay-PRA-2011, Safronova-PRL-2011}.
The contribution of the BBR shift to the clock frequency uncertainty
can be easily suppressed without reducing the ambient temperature like the cryogenic clocks
\cite{Ushijima-nphoton-2015, Huang-PRAppl-2022}, if the $\Delta\alpha(0)$ value can be determined more precisely.
The other leading uncertainties
in the Al$^+$ clock frequency measurement arise from the excess micromotion and secular motion of the trapped ion and the quadratic Zeeman shift
that are steadily suppressed up to two orders of magnitude over the past few years \cite{Rosenband-Science-2008, Chou-PRL-2010, Brewer-PRL-2019}.

Given the ever-improving precision in the Al$^+$ ion clock and there are constant efforts to bring its systematic effects down, it is the time to
look deeply into the secondary contributors to the clock uncertainty that were neglected earlier on the basis of their small magnitudes
\cite{Rosenband-Science-2008, Chou-PRL-2010, Brewer-PRL-2019}. In such a scenario, a more refined assessment of the systematic uncertainty by
taking into account the immediate higher-order physical effects is necessary. In view of this, it is imperitive to analyze the neglected
contributions to the BBR shift in which only the leading-order contributions from the electric dipole (E1) component are taken into account in
the previous theoretical studies \cite{Kallay-PRA-2011, Safronova-PRL-2011}. The shifts in the energy levels due to the E1 component of the BBR have
analogy with the Stark shifts. The next dominant contributions to the BBR that can cause energy shifts in the atomic systems can come from the
magnetic dipole (M1) component. The energy shifts due to the M1 component of the BBR lead to the Zeeman (BBRz) shifts and they were found to be of the order
of $10^{-19}$ level in some of the previously analyzed clock systems \cite{Porsev-PRA-2006, Arora-PRA-2012, Sahoo-PramanaJP-2014,Kajita-JPSJ-2019}.
Therefore, it would be interesting to find out its contribution to the Al$^+$ clock frequency, which is not yet investigated.

In this paper, we estimate the BBRz shift of the Al$^+$ clock frequency by integrating the differential dynamic polarizabilities of the clock states
along the Planck's distribution of the BBR magnetic field. We show that the fine-structure intramanifold resonance can highly suppress the BBRz shift.
It is expected that contributions from the electric quadrupole (E2) component to the BBR shift much smaller than the E1 and M1 contributions
\cite{Porsev-PRA-2006, Arora-PRA-2012}, so we are not going to discuss the E2 contributions here. The earlier description of the BBR theory had focused mainly on the
contributions from the E1 component \cite{Farley-PRA-1981}. With the demand for their precise estimates, the theory was extended later to estimate the
multipolar contributions to the BBR shifts \cite{Porsev-PRA-2006, Arora-PRA-2012, Sahoo-PramanaJP-2014}. For determining these contributions conveniently,
simple formulas in terms of scalar polarizabilities and temperature were given and they have been used extensively in the literature to estimate the BBR
shifts of numerous clock candidates. However, these formulas were derieved based on certain assumption such as the energy levels of the atomic systems
are well separated. Thus, the derivation of the BBR shift formula for the E1 component can be safely used. As pointed out by Porsev $et~al.$
\cite{Porsev-PRA-2006}, one has to be a little careful while estimating BBR shift due to other electromagnetic components for an energy level
that has the fine-structure manifold. In this case the separation between the energy levels can be comparable to the characteristic wave number of the BBR spectra.
The sensitivity of the BBRz shift to the energy-level separation has not been demonstrated quantitatively in any of the clock candidates yet. Since the excited state $3s3p~^3P_0$
of the clock transition in Al$^+$ has several fine-structure partners, it is anticipated that they would contribute significantly to the BBRz shift
of the clock frequency. We intend to investigate their roles in the BBRz shift of the clock frequency of Al$^+$, which requires accurate determination
of both the static and dynamic M1 polarizabilities of the clock states. We have employed the combined configuration interaction (CI) method and the
second-order many-body perturbation theory (MBPT method) as well as the multiconfiguration Dirac-Hartree-Fock (MCDHF) method to calculate the matrix
elements of the M1 operator for this purpose. After verifying the results for Al$^+$, we also analyze the BBRz shifts in other optical clock candidates
and find that the fractional BBRz shifts in all these clock transitions are at or below 10$^{-20}$ level due to highly suppressed contributions of the
fine-structure intramanifold resonances.

\section{\label{sec:theory}Theory}

The interaction potential seen by an electron of an atomic system in the presence of propagating electromagnetic field (in the Coulomb gauge)
is given by
\begin{eqnarray}\label{eq:V}
V(r, \omega) &=& - c \bm{\alpha} \cdot \text{\bf A}(r, \omega) \nonumber \\
             &=& - c (\bm{\alpha} \cdot \boldmath{\hat \epsilon}) \exp{(i {\bf k} \cdot \text{\bf r})},
\end{eqnarray}
where $\bm{\alpha}$ is the Dirac matrix in operator form, $\omega$ is the angular frequency of the field, and ${\bf k} = k \hat k$ and
$\boldmath{\hat \epsilon}$ are its wave vector and polarization direction, respectively.

Due to $V(r, \omega)$, one can observe change in the energy levels in the atomic system from the values obtained using the atomic Hamiltonian $H_{\textrm{at}}$.
For weak potential $V(r, \omega)$, the change in energy levels can be determined in the perturbative analysis. On the otherhand, the $V(r, \omega)$
potential can be expanded in terms of the electric and magnetic multipoles as shown below. In the BBR, this potential is isotropic because of which
all the odd-order of perturbation will vanish. Owing to the same reason, contributions to the energy shifts from the vector and tensor components
from the even-order of perturbation will also be zero. Thus, contributions arising from the scalar component of the second-order perturbation will
give the dominant contributions to the energy shifts due to the BBR.

The second-order energy shift of an atomic state $\vert \Psi_n \rangle$ with energy $E_n=\omega_n$ (in a.u.: $\hbar = e = m_e = 4\pi\epsilon_0 = 1$) can be given by
\begin{eqnarray}\label{eq:En}
\delta E_n(\omega ) = - \frac{1}{2} \sum_{n'} |V(r, \omega)|^2
\frac{\omega_{n'n}}{\omega_{n'n}^2 -\omega^2} ,
\end{eqnarray}
where $\omega_{n'n}=\omega_{n'}-\omega_n$ is the angular frequency difference between the $\vert \Psi_n \rangle$ state and an intermediate state
$\vert \Psi_{n'} \rangle$. In terms of the traditional multipole moment ($Q_{LM}^{\lambda}$) expansion, we can express
\begin{eqnarray}\label{eq:Q}
(\bm{\alpha} \cdot \boldmath{\hat \epsilon}) \exp{(i {\bf k} \cdot {\bf r})} &=&
- \sum_{L M} \frac{ k^L i^{L+1+\lambda}}{(2L+1)!!} [\text{\bf Y}_{LM}^{\lambda}(\hat{k}) \cdot \boldmath{\hat \epsilon} ] \nonumber \\
&& \sqrt{\frac{4 \pi (2L+1)(L+1)}{L}} Q_{LM}^{\lambda},
\end{eqnarray}
where $\lambda=1$ and $\lambda=0$ represents the electric and magnetic multipoles, respectively. Detailed derivation on these formulas can be found
from Refs. \cite{Porsev-PRA-2006, Arora-PRA-2012, Sahoo-PramanaJP-2014}. In the present work, our interest is limited to probe only the contribution
due to the M1 component. Thus, substituting $\lambda=0$ and $k=1$, we can get the BBRz shift expression of an atomic energy level as
\begin{eqnarray}
\delta E_{\textrm{bbrz}}(T) & = & - \frac{4\alpha^5}{3(2J+1)\pi} \int_0^\infty \sum_{n'} | \langle n || \bm{\mu} || n' \rangle |^2  \nonumber \\
& & \times \frac{\omega_{n'n}}{\omega_{n'n}^2 - \omega^2}
 \frac{\omega^3}{\textrm{exp}\left( \frac{\omega}{k_BT} \right) - 1} \textrm{d}\omega
\label{eq:BBRZ-dynamic-a} \\
  & = & - \frac{2\alpha^5}{\pi} \int_0^\infty \beta(\omega)
\frac{\omega^3}{\textrm{exp}\left( \frac{\omega}{k_BT} \right) - 1} \textrm{d}\omega  ,
\label{eq:BBRZ-dynamic-b}
\end{eqnarray}
where $\bm{\mu}$ denotes the M1 operator, double bar denotes the reduced matrix element (RME), and $\beta(\omega$) is known as the dynamic M1 polarizability, which is defined as
\begin{eqnarray}\label{eq:beta}
\beta ( \omega ) & = & \frac{2}{3(2J+1)} \sum_{n'} |\langle n || \bm{\mu} || n' \rangle|^2 \frac{\omega_{n'n}}{\omega_{n'n}^2 - \omega^2} .
\end{eqnarray}
To simplify the above integral conveniently, the
$x=\frac{\omega}{k_BT}$ and $y_{n'n}=\frac{\omega_{n'n}}{k_BT}$ variables are introduced. It gives us
\begin{equation}\label{eq:BBRZ-dynamic-c}
\delta E_{\textrm{bbrz}}(T) =
- \frac{4\alpha^5(k_BT)^3}{3(2J+1)\pi} \sum_{n'} | \langle n || \bm{\mu} || n' \rangle |^2 f(y_{n'n}) ,
\end{equation}
where
\begin{equation}\label{eq:f}
f(y_{n'n}) = \int_0^\infty \frac{y_{n'n}}{y_{n'n}^2 - x^2} \frac{x^3}{e^x - 1} \textrm{d}x.
\end{equation}
The function $f(y_{n'n})$ is similar to the function $F(y)$ of Ref. \cite{Farley-PRA-1981} for the E1 BBR shift and of Refs.
\cite{Porsev-PRA-2006, Arora-PRA-2012} for the general higher multipoles. For small and large values of $y_{n'n}$, this integral function
can be simplified (see Appendix \ref{sec:appendixA}) as
\begin{numcases}
{f(y_{n'n})\thickapprox} -\frac{\pi^2 y_{n'n}}{6},    & $|y_{n'n}|\ll1$,  \label{eq:f-a}  \\
\frac{\pi^4}{15y_{n'n}} + \frac{8\pi^6}{63y_{n'n}^3}, & $|y_{n'n}|\gg1$.  \label{eq:f-b}
\end{numcases}

In Eq. (\ref{eq:f-b}), we can drop the second term due to its negligible contribution as $|y_{n'n}| \gg 1$. This gives us
\begin{eqnarray}\label{eq:BBRZ-static-a}
\delta E_{\textrm{bbrz}}(T) &=& -\frac{1}{2} \frac{4\pi^3\alpha^5}{15} (k_BT)^4 \beta(0).
\end{eqnarray}
With respect to the room temperature, a general expression can be given as
\begin{eqnarray}\label{eq:BBRZ-static-b}
\delta E_{\textrm{bbrz}}(T) &=& -\frac{1}{2}(2.775 \times 10^{-6} \rm{Tesla})^2\left[\frac{\rm{T(K)}}{300}\right]^4 \beta (0).  \ \ \ \
\end{eqnarray}
Evaluation of the integral in the second term of Eq. (\ref{eq:f-b}) can be complicated and its contribution can be included as a small dynamic correction
if required.

Similarly, after simplifying Eq. (\ref{eq:f}) by Eq. (\ref{eq:f-a}) for $|y_{n'n}|\ll1$ it can be shown that $\delta E_{\textrm{bbrs}} \propto T^2$. However, there is a
singularity in Eq. (\ref{eq:f}) for $|y_{n'n}| \sim 1$ ($\omega_{n'n} \sim k_BT$). Thus, the BBRz shift can be strongly influenced for the values $|y_{n'n}| \sim 1$. Such condition would arise around the resonance line of the fine-structure manifold. The general perception is that
such condition would enhance the BBRz shift, but is found to be otherway around in the present study. To demonstrate this fact, we evaluate the exact
integral given by the original expressions Eqs. (\ref{eq:BBRZ-dynamic-a}) and (\ref{eq:BBRZ-dynamic-b}) in a number of optical clock transitions including the clock transition in Al$^+$. Numerical
determination of this expression also demands for calculation of the dynamic M1 polarizabilities for a wide range of $\omega$.

\section{\label{sec:calculations}Computational Methods}

To accurately determine the static and dynamic M1 polarizabilites of the clock states of the considered atomic systems in the frequency range of
interest, we calculate the M1 RMEs of the resonant transitions. Since the excitation energies between the fine-structure splitting are very small,
the M1 amplitude between the fine-structure levels will give almost all the contributions to the M1 polarizabilitis. So we focus specifically on
the accuracy of the evaluated RMEs between the fine-structure partners of the clock states for our analysis. We have employed two independent
relativistic many-body methods of atomic structure, the CI+MBPT \cite{Dzuba-PRA-1996, Kozlov-CPC-2015} and MCDHF \cite{Fischer-JPB-2016,
Fischer-CPC-2019} methods. These two methods have distinct features to incorporate the electron correlations in the calculations. Therefore,
comparison of the results of RMEs from both the methods can test accuracy of the calculations. Moreover, we had already applied these methods to
study the dynamic E1 polarizabilities of Al$^+$ \cite{Wei-CPB-2022} and Yb \cite{Tang-JPB-2018, Tang-PRA-2023} accurately.
So the potential of these methods to study
atomic properties in the similar systems have been verified. Below, we outline these methods briefly for completeness and easily understanding
of the reported results.

\subsection{\label{sec:CIMBPT}CI+MBPT method}

The hybrid CI+MBPT method combines the advantages of both the CI and MBPT methods, where the CI method explicitly accounts for the correlations between
the strongly interacting valence electrons (VV) in the presence of a ``frozen" core, and the MBPT method includes all valence-core (VC) and core-core
(CC) correlations into an effective Hamiltonian \cite{Dzuba-PRA-1996}. For the calculations of Al$^+$  ($N=Z-1=12$), the electron configurations
of the atomic states are treated as structures with two valence electrons above a closed-shell core [$1s^22s^22p^6$].

The atomic state function (ASF) of a two-electron valence state in the CI method is formed as a linear combination of a set of configuration state
functions (CSFs) with a given angular momentum $J$ and parity $P$, namely
\begin{equation}\label{eq:ASF}
\Psi(\gamma PJ) = \sum_{i=1}^M c_{i} \Phi(\gamma_{i}PJ) ,
\end{equation}
where $\Phi(\gamma_{i}PJ)$ are the CSFs in the CI model space with dimension $M$, $\{c_{i}\}$ are the expansion coefficients and $\gamma_{i}$ are the
unspecified quantum numbers. In the CI$+$MBPT method, they are generated by all possible single (S) and double (D) excitations of the two valence electrons in
a given basis set. The CSFs are further defined using the Slater determinants of the single particle Dirac orbitals given by
\begin{equation}\label{eq:Orbital}
\phi_{n \kappa m}(r,\theta,\varphi) = \frac{1}{r} \left ( \begin{matrix} P_{n \kappa}(r) & \Omega_{\kappa m}(\theta,\varphi) \cr
\textrm{i}Q_{n \kappa}(r) & \Omega_{-\kappa m}(\theta,\varphi) \cr \end{matrix} \right ) ,
\end{equation}
where the radial wave function $P_{n \kappa}(r)$ and $Q_{n \kappa}(r)$ are numerically defined on a grid, $\Omega_{\kappa m}(\theta,\varphi)$ is
the two-component spherical spinor, $n$ is the principal quantum number, $\kappa=\pm j+1/2$ (for $j=l\mp 1/2$) is the relativistic quantum number
and $m$ is the magnetic quantum number associated with the $z$-projection of total orbital angular momentum $j$. The single particle orbitals in our
calculations include the $1s$-$29s$, $2p$-$29p$, $3d$-$29d$, and $4f$-$29f$ orbitals, where the core, $3s$, $4s$, $3p$, $4p$, $3d$, $4d$ and $4f$
orbitals are obtained by solving the Dirac-Hartree-Fock (DHF) equations in the $V^{N-2}$ approximation, while the remaining orbitals are constructed
using a recurrent procedure \cite{Kozlov-CPC-2015}.

A valence state $\Psi_n$ of Al$^+$ is determined by solving the eigenfunction
\begin{equation}\label{eq:eigenfunc}
H \Psi_n = E_n \Psi_n.
\end{equation}
The CI Hamiltonian describing the valence correlation has the form (in a.u.)
\begin{eqnarray}\label{eq:H-CI}
H_{\textrm{CI}} & = &
\sum_{i=1}^2 \left[ c \bm{\alpha}_i \cdot \bm{p}_i + (\beta_i-1)  c^2 + V_{\textrm{nuc}}(r_i) + V^{N-2}(r_i) \right] \nonumber \\
& & + \frac{1}{2} \sum_{i,j=1}^2 V_{\textrm{C}}(r_{i,j}) + H_{\textrm{core}},
\end{eqnarray}
where $c$ is the speed of light, $\bm{\alpha}$ and $\beta$ are the Dirac matrices, $\bm{p}$ is the momentum operator, $V_{\textrm{nuc}}(r)$ is the nuclear potential,
$V^{N-2}(r_i)$ is the DHF potential created by the core electrons,
$V_{\textrm{C}}(r_{1,2})$ denotes the Coulomb potential with interelectronic distance $r_{1,2}$, and $H_{\textrm{core}}$ denotes the energy of the frozen core which usually cancel out in the evaluation of the excitation energies or transition matrix elements.

In the hybrid CI$+$MBPT approach, the correlation effects from the frozen core electrons are accounted through the second-order MBPT method by
defining two correlation operators $\Sigma_1$ and $\Sigma_2$ along with the dominant correlation effects due to the valence electrons through the
CI method. Thus, the net effective Hamiltonian in this hybrid approach is given by
\begin{eqnarray}\label{eq:H-CIMBPT}
H_{\textrm{eff}} =
H_{\textrm{CI}} + \Sigma_1(r_1) + \Sigma_1(r_2) + \Sigma_2(r_1,r_2) ,~~
\end{eqnarray}
where $\Sigma_1$ is the single-electron correlation operator that accounts for the correlation interactions through one-hole--one-particle excitations
and $\Sigma_2$ is the two-electron correlation operator that includes the correlation interactions via two-hole--two-particle excitations due to the
screened Coulomb potential between the valence and the core electrons \cite{Dzuba-PRA-1996}.

\begin{table*}[!hbtp]
\caption{Excitation energies (EE) and M1 RMEs (absolute values) of the resonant transitions associated with the $3s^2~^1S_0$ and $3s3p~^3P_0$ clock
states of Al$^+$, obtained from the CI, CI$+$MBPT, and MCDHF methods. The experimental values of EEs taken from the NIST database are listed for
comparison, whereas the calculated results are given as the corresponding differences of these calculations from the experiments.
``$\delta_{\textrm{Breit+QED}}$" denotes the contributions of the Breit interaction and QED effects to the M1 RMEs. The recommended values of the
M1 RMEs are given in the last column labeled as ``recommended". \label{tab:transition}}
{\footnotesize
\begin{tabular*}{\textwidth}{@{\extracolsep{\fill}}l rrcc c rrrrr@{}}\hline\hline
Transition	&			\multicolumn{4}{c}{ EE (in cm$^{-1}$) }		&	&	\multicolumn{5}{c}{ M1 RME (in a.u.: $\mu_{\textrm{B}} = e\hbar/2m_e$) }							\\ \cline{2-5} \cline{7-11}
	&	Experiment	&	$\Delta_{\textrm{CI}}$	&	$\Delta_{\textrm{CI+MBPT}}$	&	$\Delta_{\textrm{MCDHF}}$	&	&	CI	&	CI+MBPT	&	MCDHF	&	$\delta_{\textrm{Breit+QED}}$	&	Recommended	\\ \hline
$3s^2~^1S_0 - 3s4s~^3S_1$	&	91274.50 	&	$-$1267	&	$-$21	&	254 	&	&	0.00008 	&	0.00009 	&	0.00006 	&	0.00000 	&	0.00008(2)	\\
$3s^2~^1S_0 - 3p^2~^3P_1$	&	94147.46 	&	$-$1488	&	$-$40	&	142 	&	&	0.00072 	&	0.00072 	&	0.00069 	&	$-$0.00004	&	0.00071(5)	\\
$3s^2~^1S_0 - 3s3d~^3D_1$	&	95551.44 	&	$-$1381	&	$-$61	&	395 	&	&	0.00003 	&	0.00003 	&	0.00001 	&	0.00000 	&	0.00002(1)	\\
$3s^2~^1S_0 - 3s5s~^3S_1$	&	120092.92 	&	$-$1502	&	$-$46	&	301 	&	&	0.00004 	&	0.00004 	&	0.00003 	&	0.00000 	&	0.00004(1)	\\
$3s^2~^1S_0 - 3s4d~^3D_1$	&	121484.25 	&	$-$1514	&	$-$62	&	349 	&	&	0.00001 	&	0.00001 	&	0.00001 	&	0.00000 	&	0.00001(1)	\\
$3s3p~^3P_0^o - 3s3p~^3P_1^o$	&	60.88 	&	1.3 	&	4.1 	&	$-$0.7	&	&	1.41419 	&	1.41418 	&	1.41418 	&	0.00000 	&	1.41418(2)	\\
$3s3p~^3P_0^o - 3s3p~^1P_1^o$	&	22458.99 	&	824 	&	104 	&	164 	&	&	0.00456 	&	0.00498 	&	0.00466 	&	$-$0.00028	&	0.00473(46)	\\
$3s3p~^3P_0^o - 3s4p~^3P_1^o$	&	68048.47 	&	$-$329	&	$-$4	&	213 	&	&	0.00062 	&	0.00064 	&	0.00060 	&	$-$0.00004	&	0.00062(6)	\\
$3s3p~^3P_0^o - 3s4p~^1P_1^o$	&	69527.53 	&	$-$210	&	10 	&	229 	&	&	0.00095 	&	0.00099 	&	0.00092 	&	$-$0.00006	&	0.00095(9)	\\
$3s3p~^3P_0^o - 3s5p~^3P_1^o$	&	88315.80 	&	$-$429	&	$-$10	&	253 	&	&	0.00029 	&	0.00030 	&	0.00029 	&	$-$0.00002	&	0.00029(3)	\\
$3s3p~^3P_0^o - 3s5p~^1P_1^o$	&	88475.99 	&	$-$409	&	$-$4	&	259 	&	&	0.00049 	&	0.00051 	&	0.00048 	&	$-$0.00003	&	0.00049(5)	\\ \hline\hline
\end{tabular*}}
\end{table*}

\subsection{\label{sec:MCDHF}\textbf{MCDHF method}}

The MCDHF method is similar in some aspects to the CI procedure in the CI+MBPT method, where the ASF $\Psi(\gamma PJ)$ is also given as an expansion of
a set of CSFs $\psi(\gamma PJ)$ as in Eq. (\ref{eq:ASF}) to capture the electron correlation effect. Here the DHF orbitals are obtained by using the
Dirac-Coulomb (DC) Hamiltonian
\begin{eqnarray}\label{eq:H-DC}
H_{\textrm{DC}} & = &
\sum_{i=1}^N \left[ c \bm{\alpha}_i \cdot \bm{p}_i + (\beta_i-1)  c^2 + V_{\textrm{nuc}}(r_i) \right] \nonumber \\
&& + \sum_{j>i=1}^{N} V_{\textrm{C}}(r_{i,j}) ,
\end{eqnarray}
where $N$ denotes the number of electrons in the system \cite{Fischer-JPB-2016}.

We choose the $3s^2$, $3p^2$, $3s4s$, $3s5s$, $3s3d$, $3s4d$ configurations
and the $3s3p$, $3s4p$, $3s5p$, $3s4f$ configurations to construct
two multi-reference (MR) model spaces for the even and odd parity states of Al$^+$, respectively,
where the orbitals with $n\leqslant2$ are considered as the core orbitals and the other orbitals are the valence orbitals.
The ASFs of the even and odd states are optimized separately.
The initial step of the MCDHF calculation with the RSCF procedure is performed to simultaneously determine all the orbitals in the MR set without any excitations.
Then the optimization of the virtual orbitals is done by virtual excitations of the electrons from the orbitals in the MR set to an active set of orbitals with layer-by-layer using the restricted active set (RAS) approach
\cite{Sturesson-CPC-2007}.
In each step when adding a new layer of orbitals in the expansions of the CSF space, only the orbitals in the new layer are optimized and all the orbitals in the lower layers are kept frozen.
We have considered both the single-electron (S) and double-electron (D) excitations from the aforementioned MR set with $n_{\textrm{max}}=5$ up to the layer with $n_{\textrm{max}}=13$ and $l_{\textrm{max}}=4$ in eight steps,
where all possible S and D excitations from the valence orbitals and S excitations from the $2p$ core shell
are included. By doing such expansions, the important valence-valence (VV) and
valance-core (VC) correlations can be well included.
We label the CSF space constructed by the MR set as ``MR" ($n_{\textrm{max}}=5$), and the CSF space expanded in each step of the SD-excitations as ``SD$_i$" ($i\leqslant8$), in which ``SD$_i$" corresponds to the SCF space expanded to the $n_{\textrm{max}}=5+i$ layer.

Based on the orbitals obtained in each MCDHF calculation, a relativistic configuration-interaction (RCI) calculation can be further performed to include the Breit interaction and the leading order quantum electrodynamic (QED) effects [the self-energy (SE) and vacuum polarization (VP) effects].
Thus, the extended Hamiltonian employed in the RCI calculation can be given by
\begin{eqnarray}\label{eq:H-full}
H_{\textrm{DCB+QED}} =
H_{\textrm{DC}} + H_{\textrm{Breit}} + H_{\textrm{SE}} + H_{\textrm{VP}}.
\end{eqnarray}
Here only the expansion coefficients $(c_i)$ are recalculated by diagonalizing the extended Hamiltonian  Eq. (\ref{eq:H-full}), but the orbitals are kept fixed \cite{Fischer-JPB-2016}.
In the RCI procedure, one can also optionally take into account some higher-order electron correlations by further expanding the SCF space, such as by adding more configurations into the MR set or considering multiple excitations based on the previous optimized orbital basis \cite{Fischer-JPB-2016}.
In the case of Al$^+$, we have considered the higher-order correlations by further including the triple (T) excitations from the aforementioned MR set to the fifth layer of the virtual orbitals, where $n_{\textrm{max}}=10$.
Note that the CSF space expanded in each T-excitation is merged with the ``SD$_8$" space in the RCI calculation, we therefore label the CSF space expanded after the $j$th step of the T-excitations as ``SD$_8$+T$_j$" ($j\leqslant5$).

Once the ASFs are obtained, the M1 RME from an initial state $\Psi(\gamma_{i}P_{i}J_{i})$ to a final state $\Psi(\gamma_{f}P_{f}J_{f})$ is determined by
\begin{eqnarray}\label{eq:transition}
\langle\Psi(\gamma_{i}P_{i}J_{i}) \|\bm{\mu}  \| \Psi(\gamma_{f}P_{f}J_{f})\rangle  \quad\quad\quad\quad\quad\quad\quad\quad     \nonumber \\
= \sum_{i,f} c_{i} c_{f} \langle\Phi(\gamma_{i}P_{i}J_{i}) \| \bm{\mu} \| \Phi(\gamma_{f}P_{f}J_{f})\rangle .
\end{eqnarray}

\section{\label{sec:results}Results and discussion}

\subsection{Al$^+$ ion}

\subsubsection{\label{sec:results-tr}Energies and matrix elements}

\begin{figure}[t]
\centering
\includegraphics[height=7.0cm,clip]{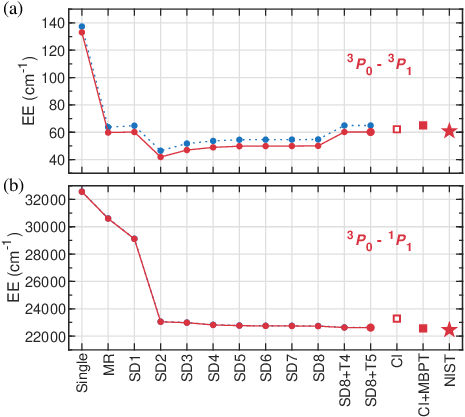}
\caption{
Deviations of the calculated excitation energies (EE) of the (a) $3s3p~^3P_0 - 3s3p~^3P_1$ and (b) $3s3p~^3P_0 - 3s3p~^1P_1$ resonant
transitions in Al$^+$ obtained using the MCDHF, CI, and CI+MBPT methods and from the NIST database. Results from the MCDHF method are given as
functions of the CSF space (see Section \ref{sec:MCDHF} for the labels). ``Single" denotes calculation using the single configuration $3s3p$ in the CSF space. The red lines and blue dotted lines indicate the calculations with and without
including the Breit and QED correlations, respectively.
\label{fig:EE}}
\end{figure}

\begin{figure}[t]
\centering
\includegraphics[height=7.0cm,clip]{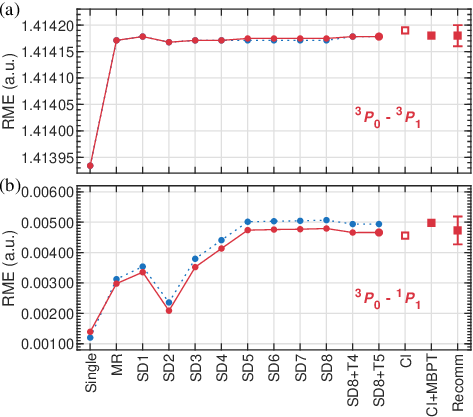}
\caption{Deviations of the calculated M1 RMEs of the (a) $3s3p~^3P_0 - 3s3p~^3P_1$ and (b) $3s3p~^3P_0 - 3s3p~^1P_1$ resonant transitions in
Al$^+$ using different basis functions with the saturated values obtained using the MCDHF, CI, and CI+MBPT methods. The recommended values with
their uncertainties are given by red squares with error bars labeled as ``Recomm". \label{fig:RME}}
\end{figure}

The excitation energies (EE) and M1 RMEs of the resonant transitions associated with the $3s^2~^1S_0$ and $3s3p~^3P_0$ clock states of
Al$^+$, obtained from the CI, CI+MBPT, and MCDHF methods, are summarized in Table \ref{tab:transition}. The experimental values for EEs taken
from the National Institute of Science and Technology (NIST) database \cite{NIST} are also listed in the same table for comparison. The convergence of
the MCDHF calculations with the increase of the configuration space is examined in detail. We plot the convergence behaviors of the EE calculations
and M1 RMEs of the two strongest M1 lines corresponding to the $3s3p~^3P_0 - 3s3p~^3P_1$ and $3s3p~^3P_0 - 3s3p~^1P_1$ transitions in
Figs. \ref{fig:EE} and \ref{fig:RME}, respectively.

In comparison with the CI calculations, the evaluated energies from the CI$+$MBPT method show significant improvement in the accuracy. The deviations
of the calculations from the experimental results are reduced by approximately one or two orders of magnitude when we include all VC and CC correlation
effects through the MBPT method. The accuracies of the EEs from the MCDHF method are generally seen to be lying in between the results from the CI and
CI$+$MBPT methods, as the MCDHF calculations have included the VC correlations involving the $2p$ core shell but ignored other VC and CC correlations.
In our test calculations, we found that the VC correlations involving the excitations from the $1s$ and $2s$ orbitals and the CC correlations from the
SD excitations of the $2p$ electrons do not improve the calculations significantly. Therefore, we do not consider more VC and CC correlations in the
MCDHF calculations, but we turn to include the triple (T) excitations into the VC correlation model. Note that the triple excitations are not included
in the CI+MBPT calculations. It was found that considering the SDT excitations can give reasonably accurate results for many levels.
As shown in Table \ref{tab:transition} and Fig. \ref{fig:EE}(a), the MCDHF method in the SDT approximation gives the most accurate fine-structure
splitting for the $3s3p~^3P_{0,1}$ states, with only $-$0.7 cm$^{-1}$ deviation from the experimental value. Although the calculated M1 RME values for the
strongest $3s3p~^3P_0 - 3s3p~^3P_1$ transition is very consistent from all the three considered methods, the improved energy kind of gives us an
assurance about the reliability in the determined M1 RME value using the MCDHF method. Another advantage of considering results from the MCDHF method
over the CI+MBPT method is that we can estimate the corrections from the Breit interaction and QED effects using the MCDHF method. We present these
corrections to the M1 RMEs in Table \ref{tab:transition}. Considering possible uncertainties from many dominant sources, we present our recommended
values of the M1 RMEs associated with the $3s^2~^1S_0$ and $3s3p~^3P_0$ clock states of Al$^+$ in the last column of Table \ref{tab:transition}.

\subsubsection{\label{sec:results-M1polariz}M1 polarizabilities}

We determine the M1 polarizabilities [see Eq. (\ref{eq:beta})] of the $3s^2~^1S_0$ and $3s3p~^3P_0$ clock states of Al$^+$ by combining the theoretical M1 RMEs
with the experimental energies in order to minimize the uncertainty in the results. The experimental energies of low-lying levels of Al$^+$ taken
from NIST database \cite{NIST} generally have uncertainties $\leqslant$ 0.06 cm$^{-1}$ \cite{Martin-JPCRD-1979}, which correspond to fractional
uncertainties of $\lesssim1\times10^{-6}$ to the EEs listed in Table \ref{tab:transition} except for the $3s3p~^3P_0 - 3s3p~^3P_1$ transition
which has a very small EE of 60.88 cm$^{-1}$. Since the $3s3p~^3P_0 - 3s3p~^3P_1$ transition has the largest M1 RME, at least three
orders of magnitude larger than other M1 transitions associated with both the clock states. Thus, an error of $\thickapprox$ 0.06 cm$^{-1}$ in
its EE of 60.88 cm$^{-1}$ can introduce a fractional error of $\thickapprox$ 0.1\% to the dynamic M1 polarizability of the $3s3p~^3P_0$ state in the BBR frequency region. Therefore, we alternatively use its EE inferred from a high-precision laser spectroscopic measurement of the $^3P_{0,1}$
fine-splitting, whose value is given in Ref. \cite{Brewer-PRA-2019} as 1.8241180(2) THz [or 60.846027(7) cm$^{-1}$] in the absence of the
hyperfine interaction. The use of high-precision energies in the evaluation of M1 polarizabilities ensure that nominal uncertainties to
these quantities can arise only from the calculated M1 RMEs.

\begin{figure*}[htbp]
\centering
\includegraphics[width=14.0cm,clip]{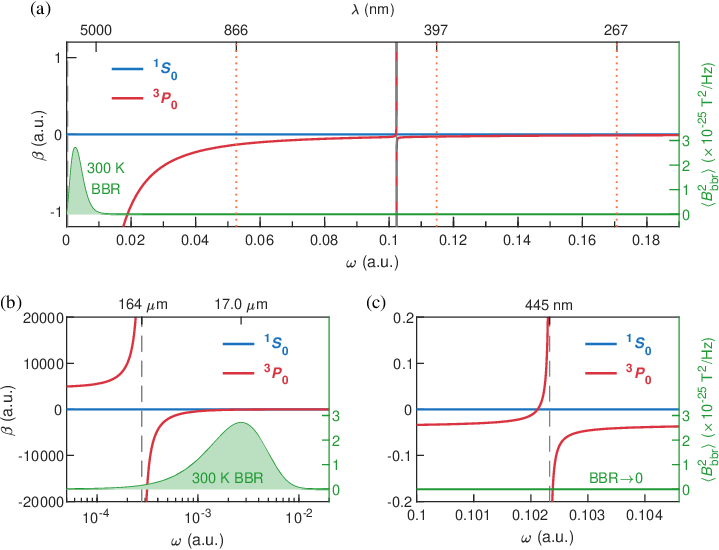}
\caption{Frequency-dependent dynamic M1 polarizabilties $\beta(\omega)$ of the $3s^2~^1S_0$ and $3s3p~^3P_0$ clock states of Al$^+$ and the Planck's distribution of the BBR magnetic field at 300 K.
The resonance behavior of the $3s3p~^3P_0$ polarizabilty is due to the 164-$\mu$m
$3s3p~^3P_0 - 3s3p~^3P_1$ and 445-nm $3s3p~^3P_0 - 3s3p~^1P_1$ resonances. The dotted lines at 866, 397, and 267 nm indicate the wavelengths of
the lasers applied during the probe procedure in the case of $^{40}$Ca$^+$-$^{27}$Al$^+$ quantum-logic clock. \label{fig:polari}}
\end{figure*}

The frequency-dependent dynamic M1 polarizabilities $\beta(\omega)$ of both the clock states of Al$^+$ are plotted in Fig. \ref{fig:polari} in the
range from $\omega=0$ (static limit) to $\omega=0.19$ a.u.. We also plot the distribution of the BBR spectrum in the same horizontal coordinate
in order to facilitate the analysis, and the positions of the laser wavelengths applied in the $^{40}$Ca$^+$-$^{27}$Al$^+$ quantum-logic clock \cite{Guggemos-NJP-2019, Hannig-RSI-2019, Ma-APB-2020, Shao-CPL-2019, Cui-EPJD-2022} are marked.
As can be seen from the figure, the $\beta(\omega)$ curve of the $3s^2~^1S_0$ state almost
overlap with the $\beta(\omega)=0$ line. The resonant lines of the $3s^2~^1S_0$ state are all located in the extreme ultraviolet region due to
their high EEs no less than 91274.50 cm$^{-1}$ [$>$ 0.41 a.u. (see Table \ref{tab:transition})]. Thus, they are not covered in the range considered
in this work. There are two resonant lines of the $3s3p~^3P_0$ state located at the 164 $\mu$m and 445 nm, both of which are transitions within the $3s3p$ configuration and are two strongest lines among all possible resonances of the clock states. The dominant
164 $\mu$m $3s3p~^3P_0 - 3s3p~^3P_1$ resonance is responsible for making large $\beta(\omega)$ values of the $3s3p~^3P_0$ state in the
low-frequency range from $\omega=0$ to the middle infrared region but with different signs on both sides of the resonance [see
Figs. \ref{fig:polari}(b)]. In the blue-detuning region to the 164-$\mu$m resonance, the $\beta(\omega)$ value is generally negative except in the
close vicinity of the 445 nm corresponding to the $3s3p~^3P_0 - 3s3p~^1P_1$ resonance [see Figs. \ref{fig:polari}(a) and \ref{fig:polari}(c)].
The other resonances in the higher-frequency region (not shown) are negligibly weak. It is worth noting that the BBR spectrum at 300 K are mainly
distributed in the blue-detuning region to the 164 $\mu$m resonance, which is in contrast with the dynamic E1 polarizabilities \cite{Wei-CPB-2022, Middelmann-PRL-2012},
where the BBR spectrum are generally distributed in the red-detuning region to all possible E1 resonances of a typical optical clock state \cite{Middelmann-PRL-2012}.

\begin{table*}[htbp]
\caption{Leading contributions to the differential static and dynamic M1 polarizabilities $\Delta\beta(\omega)$ [in a.u.: $\pi \epsilon_0 a_0^3
(c \alpha)^2$] of the $3s3p~^1S_0$ and $3s3p~^3P_0$ clock states of Al$^+$ at the selected wavelengths. ``17.0 $\mu$m" is the wavelength for the
peak intensity of the BBR at 300 K, whereas ``114 and 5.78 $\mu$m" are two wavelengths for the 10\%-peak intensity. ``866, 397, and 267 nm" are the
laser wavelengths in the $^{40}$Ca$^+$-$^{27}$Al$^+$ quantum-logic clock. \label{tab:M1polariz}}
{\footnotesize
\begin{tabular*}{\textwidth}{@{\extracolsep{\fill}}l rrrrrrr@{}}\hline\hline
Contribution	&	Static	&	114 $\mu$m	&	17.0 $\mu$m	&	5.78 $\mu$m	&	866 nm  	&	397 nm	&	267 nm	\\ \hline
$3s3p~^3P_0 - 3s3p~^3P_1$	&	4809.2(5)	&	$-$4474.3(5)	&	$-$52.001(5)	&	$-$5.9521(6)	&	$-$0.13353(1)	&	$-$0.028062(3)	&	$-$0.012693(1)	\\
$3s3p~^3P_0 - 3s3p~^1P_1$	&	0.00015(3)	&	0.00015(3)	&	0.00015(3)	&	0.00015(3)	&	0.00020(4)	&	$-$0.00057(12)	&	$-$0.00008(2)	\\
Total [$\Delta\beta(\omega) \approx \beta_{^3P_0}(\omega)$]	&	4809.2(5)	&	$-$4474.3(5)	&	$-$52.000(5)	&	$-$5.9520(6)	&	$-$0.13333(5)	&	$-$0.02862(12)	&	$-$0.01277(2)	\\ \hline\hline
\end{tabular*}}
\end{table*}

For estimating the shift due to the isotropic magnetic field of a clock transition, we are interested in the determination of the differential
polarizabilities, $\Delta\beta(\omega) = \beta_{^3P_0}(\omega) - \beta_{^1S_0}(\omega)$, at different frequencies. Numerical results show that the $\beta_{^1S_0}(\omega)$ values in the whole frequency range of interest are even smaller than the uncertainties of $\beta_{^3P_0}(\omega)$, such that
we have $\Delta\beta(\omega) \thickapprox \beta_{^3P_0}(\omega) \gg \beta_{^1S_0}(\omega)$. The contribution of the $3s3p~^3P_0 - 3s3p~^3P_1$
resonance can dominate in the evaluation of $\Delta\beta(\omega)$ [or $\beta_{^3P_0}(\omega)$] by almost 100\% in the low-frequency range
($\gtrsim$ 1000 nm) and by over 97\% in the visible region except in the vicinity of 445-nm resonance, whereas the residual fractional
contributions are almost 100\% from the $3s3p~^3P_0 - 3s3p~^1P_1$ resonance and the contributions of all other resonances are negligibly small.
We summarize the leading contributions from individual transitions to $\Delta\beta(\omega)$ at some selected wavelengths in Table
\ref{tab:M1polariz}. The intensity of the BBR at temperature 300 K is larger than 10\% of the peak intensity in the range from 114 $\mu$m to 5.78 $\mu$m, and close to zero in the visible region. On the other hand, the $\Delta\beta(\omega)$ values in the visible region are a few orders of magnitude smaller than that in the main BBR region. These values can also be used to estimate the laser-induced Zeeman shifts in Al$^+$ clocks.
Taking the $^{40}$Ca$^+$-$^{27}$Al$^+$ quantum-logic clock \cite{Guggemos-NJP-2019, Hannig-RSI-2019, Ma-APB-2020, Shao-CPL-2019, Cui-EPJD-2022} as an example, we used the $\Delta\beta(\omega)$ values at laser wavelengths 866, 397, and 267 nm to estimate first the laser-induced Zeeman shifts in our $^{40}$Ca$^+$-$^{27}$Al$^+$ clocks at APM \cite{Shao-CPL-2019, Cui-EPJD-2022} and found these shifts to be less than 1 $\mu$Hz.

To verify reliability of our calculations, we use the differential static M1 polarizabiliy $\Delta\beta(0)$ to evaluate the quadratic Zeeman shift coefficient ($C_2$), which
can be given by $C_2 = -\frac{\Delta\beta(0)}{2}$. Our computational result is $C_2 = -71.591(7)$ Hz/mT$^2$. This value agrees rather good
with two previous measurements as $-$71.988(48) \cite{Rosenband-Science-2008} and $-$71.944(24) Hz/mT$^2$ \cite{Brewer-PRA-2019}, despite the
discrepancy ($\approx0.5$ \%) being over the uncertainties. However, it is worth noting that a theoretical value $-71.927$ Hz/mT$^2$ calculated in
a very simple nonrelativistic limit in Ref. \cite{Brewer-PRA-2019} is consistent better with the measurements. In their work, only the
nonrelativistic value of M1 RME of the $3s3p~^3P_0 - 3s3p~^3P_1$ transition was used to calculate the $C_2$ coefficient. In fact, the
nonrelativistic M1 RME value [$\sqrt{2}(1+2a)\mu_B$ = 1.41749 a.u.] is found to be larger than all the relativistic values given in Table
\ref{tab:transition} and Fig. \ref{fig:RME} (b) by approximately 0.25\%. The underlying reason for the discrepancy of the experimental values from the
high-precision relativistic calculations is worth expolring by carrying out further studies using other relativistic many-body methods like
relativistic coupled-cluster theory or including higher-order relativistic effects. We note that the role of negative-energy states on the M1 and
E2 polarizabilities have been discussed in the literature \cite{Wu-arxiv-2023-1, Wu-arxiv-2023-2,
Porsev-arxiv-2023} and these contributions are found to be important only in the visible region where the multipolar polarizabilities are generally very small.
Nonetheless, the M1 polarizability values determined in this work for the BBR region are accurate enough to estimate the BBRz shift of the clock transition of Al$^+$.

\subsubsection{\label{sec:results-BBRZ}BBR-induced Zeeman shift}

\begin{figure}[t]
\centering
\includegraphics[width=8.0cm,clip]{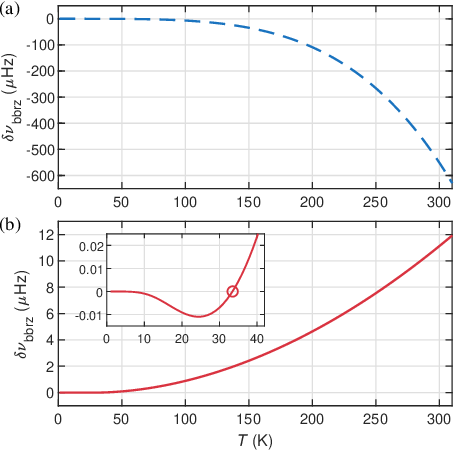}
\caption{
The BBRz shift $\delta\nu_{bbrz}(T)$ in the Al$^+$ optical clock as a function of the ambient temperature $T$, estimated using (a) the static
M1 polarizability limit and (b) considering dynamic M1 polarizabilities over all possible frequency range. The red circle in the zoom-in region of
the panel (b) denotes the zero-crossing point at the critical temperature $T_c$.
\label{fig:BBRZ}}
\end{figure}

As aforementioned, the BBR spectrum are mainly distributed in the blue-detuning region to the 164 $\mu$m $3s3p~^3P_0 - 3s3p~^3P_1$ resonance,
where the dynamic M1 polarizabilities are smaller than the static value and have an opposite sign. This has suggested the invalidation of the
static limit for the differential M1 polarizability to estimate the BBRz shift as discussed in section \ref{sec:theory}. To understand how the BBRz shift fails to achieve this limit in the Al$^+$ clock frequency,
we discuss the numerical results from both the static and dynamic M1 polarizabilities here. Using the $\Delta\beta(0)$ value listed in Table \ref{tab:M1polariz}, the BBRz shift is calculated to be $-$551 $\mu$Hz at 300 K. The temperature dependence of the BBRz shift in the static polarizability limit is given by the blue-dashed curve in Fig. \ref{fig:BBRZ} (a). Our final estimation of the
BBRz shift is obtained by using the all-frequency dynamic polarizability integral given by Eq. (\ref{eq:BBRZ-dynamic-b}) and the result is found to be 11.1 $\mu$Hz at 300 K. The temperature dependence of the BBRz shift from the dynamic polarizability integral approach is shown by the red curve in Fig. \ref{fig:BBRZ} (b).
It is obvious that there are large contradictions between
these two results. The curve in Fig. \ref{fig:BBRZ} (a) following the static-limit formula Eq.~(\ref{eq:BBRZ-static-a}) is a simple quartic function curve,
whereas the curve in Fig. \ref{fig:BBRZ} (b) drawn from the dynamic polarizability integral formula Eq. (\ref{eq:BBRZ-dynamic-b}) bends in the opposite direction.
The BBRz shift at 300 K estimated from the dynamic integral is only about 2\% of that from the static limit with an opposite sign. This is mainly
because the absolute value of $\beta_{^3P_0}(\omega)$ descends fast with the increase of the frequency in the blue-detuning region to the
$3s3p~^3P_0 - 3s3p~^3P_1$ resonance [see Fig. \ref{fig:polari}(b)]. Since the $3s3p~^3P_0 - 3s3p~^3P_1$ resonance dominates the contribution of
$\Delta\beta(\omega)$ [$\approx \beta_{^3P_0}(\omega)$] in the BBR region, it can be inferred from Eq.~(\ref{eq:beta}) that $|\Delta\beta(\omega)|
\lesssim \Delta\beta(0)$ in the region $\omega \gtrsim \sqrt{2} \omega_{^3P_0 - ^3P_1} = 86$ cm$^{-1}$. Moreover, the wave number for the peak intensity of the BBR at 300 K is approximately 588 cm$^{-1}$, which is about ten times larger than the wave number of
the $^3P_0 - ^3P_1$ transition ($\omega_{^3P_0 - ^3P_1}$ = 61 cm$^{-1}$). As a consequence, the BBRz shift is highly suppressed by blue-detuning to the $3s3p~^3P_0 - 3s3p~^3P_1$ resonance for which
$|\Delta\beta(\omega)| \approx \Delta\beta(0) (\frac{\omega_{^3P_0 - ^3P_1}}{\omega})^2 \approx \frac{\Delta\beta(0)}{100}$.

Apart from the main components of the BBR spectrum that are distributed in the blue-detuning region to the $3s3p~^3P_0 - 3s3p~^3P_1$ resonance,
there are also small weak components distributed in the red-detuning region [see Fig. \ref{fig:polari} (b)]. They lead to a small cancellation of
the contributions to the BBRz shift due to opposite values of $\Delta\beta(\omega)$ on both sides of the resonance. When the temperature
declines, the intensity of the BBR decreases and meanwhile their distributions are red-shifted, such that there are more cancellated proportions in the contributions to the BBRz shift. In a more detailed view of the curve in Fig. \ref{fig:BBRZ} (b), we can find a reversal behavior along with the temperature
change in a low-temperature range, which is caused by variations of the cancellations between the contributions from both sides of the resonant line when the resonant line across the main BBR spectrum.
The reversal
behavior inevitably results in a vanishing BBRz shift at a zero-crossing point, which is predicted at a critical temperature $T_c = 34$ K for
the Al$^+$ clock. At this point, the distribution of the BBR spectrum will be divided by the resonant line approximately in halves.
From the Wien's displacement law, we can infer that the peak intensity of the BBR at 34 K is actually at $\omega_{\textrm{peak}} = 66$ cm$^{-1}$, which is very close to
the resonant frequency as $\omega_{\textrm{reson}} = 61$ cm$^{-1}$. When the temperature drops below $T_c$, contributions from the red-detuned
components go beyond that from the blue-detuned components. When the temperature further declines, the BBR spectrum can be even shifted to the
red-detuning region, and the curve describing the temperature dependence of the BBRz shift in Fig. \ref{fig:BBRZ}(b) can approach the curve
in Fig. \ref{fig:BBRZ}(a), which means that the static polarizability limit can be applicable in the low-temperature condition.

\begin{figure}[t]
\centering
\includegraphics[width=8.04cm,clip]{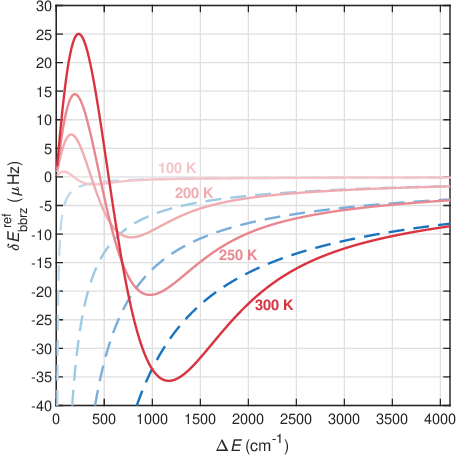}
\caption{Reference BBRz shift $\delta E_{\textrm{bbrz}}^{\textrm{ref}}(\Delta E,T)$
as a function of both the single resonant energy $\Delta E$ in different clock state and the ambient temperature $T$, estimated using the static M1 polarizability limit (blue dashed) and the dynamic M1 polarizability integral (red) approaches, respectively,
for $T =$ 300, 250, 200, 100 K.
$\Delta E$ is generally corresponds to a fine-structure splitting in the clock state.
A value of $\delta E_{\textrm{bbrz}}^{\textrm{ref}}(\Delta E,T)$ extracted from the red curve for the temperature $T$ at the resonant energy $\Delta E$ associated with the clock state can be used in Eq. (\ref{eq:BBRZ-other}) to estimate the contribution of the corresponding M1 resonant line to the BBRz shift of this clock state at $T$.
\label{fig:BBRZ-reson}}
\end{figure}

\subsection{\label{sec:results-other}Extending to other clock candidates}

Since the reversal behavior of the temperature dependence of the BBRz shift in the clock transition of Al$^+$ is predicted in this study, it is
obvious to anticipate that such behaviors can exist in other clock candidates. This could occur at different critical temperature depending on
the magnitudes of the fine-structure splitting ($\Delta E$) in the clock states of those candidates. Below we focus on discussing the dependence
of the BBRz shifts on $\Delta E$ for different clock transitions but at a given temperature, particularly at 300 K.
To simplify the analysis, we start with the simple case that is similar to the clock transition in Al$^+$,
where the differential dynamic polarizability $\Delta\beta(\omega)$ in the BBR region is only dominated by a single fine-structure intramanifold resonance associated with only one of its clock states.
In such case, the BBRz shift in the clock transition frequency [$\delta \nu_{\textrm{bbrz}}(T)$] is determined only by the energy shift of one clock state [$\delta E_{\textrm{bbrz}}(T)$] and contributed from a single fine-structure splitting,
thus a single term of the summation in Eq. (\ref{eq:BBRZ-dynamic-a}) is required in the estimation.
Using the corresponding values of the resonant angular frequency ($\omega_{n'n} = \Delta E$) and its M1 RME ($\langle n || \bm{\mu} || n' \rangle$) with a single term in Eq. (\ref{eq:BBRZ-dynamic-a}), the BBRz shift of this clock state is given by
\begin{eqnarray}\label{eq:BBRZ-dynamic-single}
\delta E_{\textrm{bbrz}}(T) & = & - \frac{4\alpha^5}{3(2J+1)\pi} \int_0^\infty  | \langle n || \bm{\mu} || n' \rangle |^2  \nonumber \\
& & \times \frac{\Delta E}{\Delta E^2 - \omega^2}
 \frac{\omega^3}{\textrm{exp}\left( \frac{\omega}{k_BT} \right) - 1} \textrm{d}\omega .
\end{eqnarray}
In different clock states, the values of $\Delta E$ would always differ largely between each other but the values of M1 RMEs can have comparable values.
Therefore, it can be useful for us to chose a unified M1 RME ($\langle n || \bm{\mu} || n' \rangle_{\textrm{ref}}$) with a unified total angular momentum ($J_{\textrm{ref}}$) as reference values in Eq. (\ref{eq:BBRZ-dynamic-single}) to observe the dependence of the BBRz shift on the magnitudes of the fine-structure splitting $\Delta E$, which
is given by a function of both $\Delta E$ and $T$ as follows
\begin{eqnarray}\label{eq:BBRZ-dynamic-ref}
\delta E_{\textrm{bbrz}}^{\textrm{ref}}(\Delta E,T) & = & - \frac{4\alpha^5}{3(2J_{\textrm{ref}}+1)\pi} \int_0^\infty  | \langle n || \bm{\mu} || n' \rangle_{\textrm{ref}} |^2  \nonumber \\
& & \times \frac{\Delta E}{\Delta E^2 - \omega^2}
 \frac{\omega^3}{\textrm{exp}\left( \frac{\omega}{k_BT} \right) - 1} \textrm{d}\omega.
\end{eqnarray}
Since the M1 RMEs of the clock states of Al$^+$ have been estimated with reasonable accuracy in this work, we conveniently use the M1 RME of the
dominant resonance $3s3p~^3P_0 - 3s3p~^3P_1$ as the reference M1 RME, which is given as $\langle n || \bm{\mu} || n' \rangle_{\textrm{ref}} = 1.41418$ a.u. in
Table \ref{tab:transition}. Substituting this value and $J_{\textrm{ref}}=0$ in Eq. (\ref{eq:BBRZ-dynamic-ref}), we calculated the
$\delta E_{\textrm{bbrz}}^{\textrm{ref}}(\Delta E,T)$ in a broad range of $\Delta E$ at 300, 250, 200, and 100 K, respectively, as shown in
Fig. \ref{fig:BBRZ-reson}.
For any clock state having specific values of $\Delta E$, $\langle n || \bm{\mu} || n' \rangle$, and $J$, its BBRz shift $\delta E_{\textrm{bbrz}}(T)$ at a given temperature $T$ obtained from Eq. (\ref{eq:BBRZ-dynamic-single}) is simply proportion to a value of $\delta E_{\textrm{bbrz}}^{\textrm{ref}}(\Delta E,T)$ given by Eq. (\ref{eq:BBRZ-dynamic-ref})
at the corresponding $\Delta E$ and $T$, which is expressed as
\begin{equation}\label{eq:BBRZ-ratio}
\frac{\delta E_{\textrm{bbrz}}(T)}{\delta E_{\textrm{bbrz}}^{\textrm{ref}}(\Delta E,T)} =
\frac{2J_{\textrm{ref}}+1}{2J+1} \left( \frac{|\langle n || \bm{\mu} || n' \rangle|}{|\langle n || \bm{\mu} || n' \rangle_{\textrm{ref}}|} \right)^2 ,
\end{equation}
Therefore, combining use the correct RME ($\langle n || \bm{\mu} || n' \rangle$) of the considered clock state and its corresponding reference value of $\delta E_{\textrm{bbrz}}^{\textrm{ref}}(\Delta E,T)$ extracted from Fig. \ref{fig:BBRZ-reson} by the correct $\Delta E$, we can fast estimate the BBRz shift for this clock state with Eq. (\ref{eq:BBRZ-ratio}),
which is given by
\begin{equation}\label{eq:BBRZ-other}
\delta E_{\textrm{bbrz}}(T) =
\frac{\delta E_{\textrm{bbrz}}^{\textrm{ref}}(\Delta E,T)}{2J+1} \left( \frac{|\langle n || \bm{\mu} || n' \rangle|}{1.41418} \right)^2 .
\end{equation}
So far we have managed the cases with only a single dominant resonance,
in which we have the BBRz frequency shift $\delta \nu_{\textrm{bbrz}} = \delta E_{\textrm{bbrz}}$.
For some complex cases, however, the number of the dominant M1
resonances in a clock state can be more than one and both clock states could have dominant contributions to the BBRz shift in the clock transition. In such cases, Eq. (\ref{eq:BBRZ-other}) can be considered as the contribution from an individual resonance in a clock state,
and the final value of the BBRz shift in the clock transition $|\textrm{g}\rangle \leftrightarrow |\textrm{e}\rangle$ can be obtained by summing over the contributions from all dominant resonances as $\delta \nu_{\textrm{bbrz}} = \sum\delta E_{\textrm{bbrz}}(\textrm{e}) - \sum\delta E_{\textrm{bbrz}}(\textrm{g})$.

To analyse the role of the fine-structure intramanifold resonance at different magnitudes of $\Delta E$, we have also presented the values of $\delta E_{\textrm{bbrz}}^{\textrm{ref}}(\Delta E,T)$ from the static polarizability limit in Fig. \ref{fig:BBRZ-reson} for comparison.
As can be seen from the figure, the shifts estimated from the static polarizability limit
can approach to that from the dynamic polarizability integral only when the fine-structure splitting $\Delta E$ is sufficiently large (approximately $\Delta E > 3500$ cm$^{-1}$
for 300 K). In this high-energy condition, the BBRz shifts gradually get smaller ($\delta \nu_{\textrm{bbrz}} \lesssim 10$ $\mu$Hz) as their resonant
lines are far off the BBR region. In the low-energy range, however, the results from the static polarizability limit are in contrast to those from
the dynamic polarizability integral. The curves derived from the dynamic polarizability integral oscillate in the low-energy range due to the resonant
lines crossing the BBR spectrum. In the range $\Delta E \lesssim 1000$ cm$^{-1}$ for 300 K, the absolute values of the BBRz shifts are highly suppressed
in comparison with the estimations from the static polarizability limit. The zero-crossing point where the BBRz shift vanishes is found at
$\Delta E_c = 546$ cm$^{-1}$ for 300 K (see Fig. \ref{fig:BBRZ-reson}). In an actual case, however, the fine-structure splitting $\Delta E$ may not be 546 cm$^{-1}$ coincidentally, but a clock state with its $\Delta E$ close to the value of $\Delta E_c$ can have a smaller BBRz shift. Note that $\Delta E_c$ as
546 cm$^{-1}$ is very close to 588 cm$^{-1}$, which is the wave number for the peak intensity of the BBR spectrum at 300 K, where a substantial
cancellation of the shift can occur between the contributions from both sides of the resonant line.

\begin{table*}[htbp]
\caption{The M1 contributions to the BBRz shifts at 300 K in a number of developing optical clocks, estimated in the nonrelativistic limit.
``$\lambda$" denotes the wavelength of the clock transitions. The excitation energies and nonrelativistic values of M1 RMEs (absolute values) of
the fine-structure intramanifold resonances associated with the clock states are given in the columns labeled as ``$\Delta E$" and ``RME",
respectively. The absolute and fractional BBRz shifts due to the M1 component are given in the last two columns labeled as ``$\delta \nu_{\textrm{bbrz}}$" and ``$\delta \nu_{\textrm{bbrz}}/\nu_0$", respectively. \label{tab:other}}
{\footnotesize
\begin{tabular*}{\linewidth}{@{\extracolsep{\fill}}llccrcrrr@{}}\hline\hline
System	&	Clock transition	&	$\lambda$ (nm)	&	Resonance	&	$\Delta E$ (cm$^{-1}$)	&	RME (a.u.)	&			\multicolumn{2}{r}{$\delta \nu_{\textrm{bbrz}}$ ($\mu$Hz)}	&	$\delta \nu_{\textrm{bbrz}} / \nu_0$	\\ \hline
Be	&	$2s^2~(^1S_0) - 2s2p~(^3P_0)$ \cite{Wu-NJP-2023}	&	455 	&	$^3P_0 - ^3P_1$	&	0.6 	&	1.417 	&		&	0.1 	&	$ 1.8\times10^{-22}$	\\
Mg	&	$3s^2~(^1S_0) - 3s3p~(^3P_0)$ \cite{Kulosa-PRL-2015}	&	458 	&	$^3P_0 - ^3P_1$	&	20 	&	1.417 	&		&	3.9 	&	$ 6.0\times10^{-21}$	\\
Al$^+$	&	$3s^2~(^1S_0) - 3s3p~(^3P_0)$ \cite{Rosenband-PRL-2007}	&	267 	&	$^3P_0 - ^3P_1$	&	61 	&	1.417 	&		&	11.2 	&	$ 1.0\times10^{-20}$	\\
Sr	&	$5s^2~(^1S_0) - 5s5p~(^3P_0)$ \cite{Takamoto-Nature-2005}	&	698 	&	$^3P_0 - ^3P_1$	&	187 	&	1.417 	&		&	24.2 	&	$ 5.6\times10^{-20}$	\\
Cd	&	$4d^{10}5s^2~(^1S_0) - 4d^{10}5s5p~(^3P_0)$ \cite{Yamaguchi-PRL-2019}	&	332 	&	$^3P_0 - ^3P_1$	&	542 	&	1.417 	&		&	0.4 	&	$ 4.3\times10^{-22}$	\\
In$^+$	&	$4d^{10}5s^2~(^1S_0) - 4d^{10}5s5p~(^3P_0)$ \cite{Ohtsubo-Hyperfine-2019}	&	237 	&	$^3P_0 - ^3P_1$	&	1075 	&	1.417 	&		&	$-$35.3	&	$-2.8\times10^{-20}$	\\
Yb	&	$4f^{14}6s^2~(^1S_0) - 4f^{14}6s6p~(^3P_0)$ \cite{Lemke-PRL-2009}	&	578 	&	$^3P_0 - ^3P_1$	&	704 	&	1.417 	&		&	$-$16.6	&	$-3.2\times10^{-20}$	\\
	&	$4f^{14}6s6p~(^3P_0) - 4f^{13}5d6s^2~(^3P_2^*)$ \cite{Safronova-PRL-2018}	&	1695 	&	$^3P_0 - ^3P_1$	&	704 	&	1.417 	&	16.6 	&	15.1 	&	$ 8.5\times10^{-20}$	\\
	&		&		&	$^3P_2^* - ^3P_1^*$	&	5668 	&	1.585 	&	$-$1.5	&		&		\\
	&	$4f^{14}6s^2~(^1S_0) - 4f^{14}6s6p~(^3P_2)$ \cite{Yamaguchi-NJP-2010, Dzuba-PRA-2018-yb}	&	507 	&	$^3P_2 - ^3P_1$	&	$-$1718	&	1.585 	&		&	6.9 	&	$ 1.2\times10^{-20}$	\\
	&	$4f^{14}6s^2~(^1S_0) - 4f^{13}5d6s^2~(^3P_2^*)$ \cite{Dzuba-PRA-2018-yb, Ishiyama-PRL-2023, Kawasaki-PRA-2023}	&	431 	&	$^3P_2^* - ^3P_1^*$	&	5668 	&	1.585 	&		&	$-$1.5	&	$-2.2\times10^{-21}$	\\
	&	$4f^{14}6s6p~(^3P_2) - 4f^{13}5d6s^2~(^3P_2^*)$ \cite{Tang-PRA-2023}	&	2875 	&	$^3P_2 - ^3P_1$	&	$-$1718	&	1.585 	&	$-$6.9	&	$-$8.4	&	$-8.0\times10^{-20}$	\\
	&		&		&	$^3P_2^* - ^3P_1^*$	&	5668 	&	1.585 	&	$-$1.5	&		&		\\
Hg	&	$4f^{14}5d^{10}6s^2~(^1S_0) - 4f^{14}5d^{10}6s6p~(^3P_0)$ \cite{McFerran-PRL-2012}	&	266 	&	$^3P_0 - ^3P_1$	&	1767 	&	1.417 	&		&	$-$26.5	&	$-2.3\times10^{-20}$	\\
Ca$^+$	&	$4s~(^2S_{1/2}) - 3d~(^2D_{5/2})$ \cite{Chwalla-PRL-2009, Huang-PRL-2016}	&	729 	&	$^2D_{5/2} - ^2D_{3/2}$	&	$-$61	&	1.553 	&		&	$-$2.2	&	$-5.4\times10^{-21}$	\\
Sr$^+$	&	$5s~(^2S_{1/2}) - 4d~(^2D_{5/2})$ \cite{Margolis-Science-2004, Steinel-PRL-2023}	&	674 	&	$^2D_{5/2} - ^2D_{3/2}$	&	$-$280	&	1.553 	&		&	$-$4.9	&	$-1.1\times10^{-20}$	\\
Ba$^+$	&	$6s~(^6S_{1/2}) - 5d~(^2D_{5/2})$ \cite{Arnold-PRL-2020}	&	1762 	&	$^2D_{5/2} - ^2D_{3/2}$	&	$-$801	&	1.553 	&		&	4.9 	&	$ 2.9\times10^{-20}$	\\
Ra$^+$	&	$7s~(^7S_{1/2}) - 6d~(^2D_{5/2})$ \cite{Holliman-PRL-2022}	&	728 	&	$^2D_{5/2} - ^2D_{3/2}$	&	$-$1659	&	1.553 	&		&	5.7 	&	$ 1.4\times10^{-20}$	\\
Yb$^+$	&	$4f^{14}6s~(^2S_{1/2}) - 4f^{14}5d~(^2D_{5/2})$ \cite{Roberts-PRA-1999}	&	411 	&	$^2D_{5/2} - ^2D_{3/2}$	&	$-$1372	&	1.553 	&		&	6.8 	&	$ 9.3\times10^{-21}$	\\
	&	$4f^{14}6s~(^2S_{1/2}) - 4f^{14}5d~(^2D_{3/2})$ \cite{Schneider-PRL-2005}	&	436 	&	$^2D_{3/2} - ^2D_{5/2}$	&	1372 	&	1.553 	&		&	$-$10.2	&	$-1.5\times10^{-20}$	\\
	&	$4f^{14}6s~(^2S_{1/2}) - 4f^{13}6s^2~(^2F_{7/2})$ \cite{Huntemann-PRL-2012, Furst-PRL-2020}	&	467 	&	$^2F_{7/2} - ^2F_{5/2}$	&	10149 	&	1.856 	&		&	$-$0.7	&	$-1.1\times10^{-21}$	\\
Lu$^+$	&	$4f^{14}6s^2~(^1S_0) - 4f^{14}5d6s~(^3D_1)$ \cite{Arnold-ncommun-2018, Zhiqiang-sciadv-2023}	&	848 	&	$^3D_1 - ^3D_2$	&	639 	&	2.126 	&		&	$-$7.7	&	$-2.2\times10^{-20}$	\\
	&	$4f^{14}6s^2~(^1S_0) - 4f^{14}6s6p~(^3D_2)$ \cite{Arnold-ncommun-2018}	&	804 	&	$^3D_2 - ^3D_1$	&	$-$639	&	2.126 	&	4.6 	&	$-$7.7	&	$-2.1\times10^{-20}$	\\
	&		&		&	$^3D_2 - ^3D_3$	&	1764 	&	2.165 	&	$-$12.4	&		&		\\
Hg$^+$	&	$4f^{14}5d^{10}6s~(^2S_{1/2}) - 4f^{14}5d^{9}6s^2~(^2D_{5/2})$ \cite{Oskay-PRL-2006}	&	282 	&	$^2D_{5/2} - ^2D_{3/2}$	&	15041 	&	1.553 	&		&	$-$0.4	&	$-4.2\times10^{-22}$	\\ \hline\hline
\end{tabular*}}
\end{table*}

Currently, several neutral atoms and singly charged ions have been widely-used to develop atomic clocks in which the $ns^2~^1S_0 - nsnp~^3P_0$
\cite{Wu-NJP-2023, Kulosa-PRL-2015, Takamoto-Nature-2005, Yamaguchi-PRL-2019, Ohtsubo-Hyperfine-2019, Lemke-PRL-2009, McFerran-PRL-2012}
and $(n+1)s~^2S_{1/2} - nd~^2D_{5/2}$ \cite{Chwalla-PRL-2009, Huang-PRL-2016, Margolis-Science-2004, Steinel-PRL-2023, Arnold-PRL-2020,
Holliman-PRL-2022, Roberts-PRA-1999, Oskay-PRL-2006} forbidden transitions are considered for precise frequency measurements. Similar to the case of
Al$^+$, the dynamic M1 polarizabilities $\Delta\beta(\omega)$ of other clock states in the BBR region are generally dominated by the fine-structure
intramanifold resonance of the clock states, whose M1 RMEs are typically orders of magnitude larger than that of other resonances. Therefore, it would
be appropriate to consider only the fine-structure intramanifold resonances for estimating the BBRz shifts of other atomic systems qualitatively. Here
Eq. (\ref{eq:BBRZ-other}) combined with the reference values from Fig. \ref{fig:BBRZ-reson} can be safely used to find the final values instead of
performing the actual integration. Specifically, we need only to find the corresponding value of $\delta E_{\textrm{bbrz}}^{\textrm{ref}}(\Delta E,T)$ in Fig.
\ref{fig:BBRZ-reson} according to the fine-structure splitting $\Delta E$ of the associated clock states, then the corrected value of the BBRz
shift can be extrapolated by using its M1 RME in Eq. (\ref{eq:BBRZ-other}).

We further use the nonrelativistic RMEs to estimate the BBRz shifts for some of the under-developing clock systems. In the nonrelativistic limit, the
M1 operator has the form $\bm{\mu}=-\mu_B(\textsl{g}_L\bm{L}+\textsl{g}_S\bm{S})$, where $\bm{L}$ and $\bm{S}$ are the total orbital and spin angular
momentum operators, and $\textsl{g}_L=1$ and $\textsl{g}_S\approx2.00232$ \cite{Fan-PRL-2023} are the orbital and spin $\textsl{g}$-factors, respectively.
The RMEs of the angular momentum operators between two states $|n\rangle$ and $|n'\rangle$ are given by \cite{Brewer-PRA-2019}
\begin{eqnarray}\label{eq:RME-L}
\langle n || \bm{L} || n' \rangle & = & \delta_{\gamma,\gamma'}\delta_{L,L'}\delta_{S,S'} (-1)^{L+S+J'+1}
\left\{
\begin{array}{ccc}
J & 1 & J' \\
L & S & L \\
\end{array}
\right\}
\nonumber \\
& \times & \sqrt{(2J+1)(2J'+1)L(L+1)(2L+1)}
\end{eqnarray}
and
\begin{eqnarray}\label{eq:RME-S}
\langle n || \bm{S} || n' \rangle & = & \delta_{\gamma,\gamma'}\delta_{L,L'}\delta_{S,S'} (-1)^{L+S+J+1}
\left\{
\begin{array}{ccc}
J & 1 & J' \\
S & L & S \\
\end{array}
\right\}
\nonumber \\
& \times & \sqrt{(2J+1)(2J'+1)S(S+1)(2S+1)} .
\end{eqnarray}
It is obvious from the above expressions that the nonrelativistic M1 RMEs are zero between different fine-structure manifolds, thus the finite M1 RMEs are induced only due to the relativistic effects.
It means that in the neutral atoms and single charged ions with weak relativistic effects, the M1 RMEs between different
fine-structure manifolds are reasonably very small and the contributions to M1 dynamic polarizabilities in the BBR region are generally dominated by the fine-structure intramanifold resonances.
Using the nonrelativistic RMEs and experimental energies $\Delta E$ from the
NIST database \cite{NIST}, we have estimated the BBRz shifts for various developing optical clocks, as summarized in Table \ref{tab:other}.
It is shown that the fractional BBRz shift in all these clock transitions are below $10^{-19}$ level. The BBRz shifts in these clock transitions
are generally determined by only one resonant line associated with a clock state in the clock transition, except for two clock transitions in
Yb ($^3P_0 - ^3P_2^*$ and $^3P_2 - ^3P_2^*$) and one clock transition in Lu$^+$ ($^1S_0 - ^3D_2^*$) which have two contributions. These shifts
mainly depend on the resonant energy $\Delta E$ as their RMEs generally have comparable values. It is found that three clock transitions have very
small BBRz shifts as $<$0.5 $\mu$Hz (at $10^{-22}$ level): $\delta \nu_{\textrm{bbrz}}^{\textrm{Be}} = 0.1$ $\mu$Hz in Be is highly suppressed by far-blue-detuning of the BBR spectrum to the M1 resonant line which has a very small $\Delta E$ as 0.6 cm$^{-1}$; in contrast, $\delta \nu_{\textrm{bbrz}}^{\textrm{Hg}^+} = -0.4$ $\mu$Hz in Hg$^+$ is highly suppressed by far-red-detuning of the BBR spectrum to the M1 resonant line which has a very large $\Delta E$ as 15041 cm$^{-1}$; whereas $\delta \nu_{\textrm{bbrz}}^{\textrm{Cd}} = 0.4$ $\mu$Hz in Cd is because its M1 resonant energy $\Delta E = 542$ cm$^{-1}$ is very close to the zero-crossing point $\Delta E_c = 546$ cm$^{-1}$ at 300 K.
It is shown that the nonrelativistic value 11.2 $\mu$Hz in Al$^+$ is consistent well with the aforementioned relativistic value 11.1 $\mu$Hz by approximately 99\%.
Further comparing the nonrelativistic RMEs with the high-precision relativistic RMEs given in other literatures such as Refs. \cite{Arora-PRA-2012,
Sahoo-PramanaJP-2014}, it can be expected that the overall uncertainty of the BBRz shifts tabulated in Table \ref{tab:other} is at the 1 \% level.

Apart from the neutral atoms and single charged ions, the BBRz shifts in the highly charged ion (HCI) clocks can also be estimated in the
nonrelativistic limit, although the uncertainty will be relatively larger (approximately at 10 \% level) due to stronger relativistic effects in these cases. To preliminarily extend the above analyses to other clock candidates which are not included in Table \ref{tab:other}, including the HCI clocks, we assume slightly higher
magnitudes of the M1 RMEs and resonant contributions in Eq. (\ref{eq:BBRZ-other}) and realize that the BBRz shifts at the room temperature in those clock candidates might be also at the $\mu$Hz level or even below,
as the reference values $\delta E_{\textrm{bbrz}}^{\textrm{ref}}(\Delta E,T)$ given in Fig. \ref{fig:BBRZ-reson}
are limited in the scale of $0\sim36$ $\mu$Hz for 300 K due to highly suppression from the fine-structure intramanifold resonances.

\section{\label{sec:conclusion}Conclusion}

In view of the unprecedent accuracy of the atomic clock frequency measurements in the laboratory, we have investigated
the roles of the black-body radiation induced Zeeman shifts (BBRz shifts) to the clock frequency measurements in the atomic clocks, especially in the singly charged alminium ion optical clock. We have employed two different relativistic many-body methods to calculate the magnetic dipole transition
matrix elements accurately, then they are combined with the experimental energies to determine dynamic magnetic dipole polarizabilities precisely
for a wide range of frequency. Using these results, we analyzed the differential BBRz energy shifts in the $^1S_0-^3P_0$ clock transition of the singly
charged aluminium ion due to the magnetic dipole contribution from the black-body radiation. From these analyses, we found that the shift is highly
suppressed due to blue-detuning of the black-body radiation spectrum to the $^3P_0-^3P_1$ fine-structure intramanifold resonance, suggesting
that the static polarizability limit conventionally used in the estimation of black-body radiation shifts break down in such case. The resonance also leads to a reversal of the BBRz shift along the change of the temperature at a very low temperature region, in which an interested zero-crossing cancellation of the shift is predicted. In addition, we have extended these studies to other clock candidates by analysing the resonant-energy dependence of the BBRz shifts for various under-developing optical clocks, and found that these shifts are of order of micro-hertz which are similarly highly suppressed by the fine-structure intramanifold resonances.
These results are useful for rigorously assessing systematic shifts in high-precision optical frequency metrology.


\begin{acknowledgments}
We thank Prof. M. G. Kozlov for helpful suggestions on the use of the CI-MBPT package.
We are grateful to Dr. K. F. Cui for useful discussions on the Al$^+$ optical clock, and Dr. B. Q. Lu for useful discussions and providing information on verifying our results on the Sr optical lattice clock.
This work was supported by the National Natural Science Foundation of China
(Grants No. 11704076, No. 11911530229, No. U1732140, No. 2021AMF01002, No. 1916321TS00103201, and No. 11934014),
the National Key Research and Development Program of China (Approved Grant No. 2020YFB1902100 and Grant No. 2017YFA0304401),
the Strategic Priority Research Program of the Chinese Academy of Sciences (Grant No. XDB21030100),
the IDP of CAS (Approved Grant No.YJKYYQ20180013).
and the Natural Science Foundation of Hubei Province (Grant Number: 2022CFA013).
B.K.S. acknowledges use of the ParamVikram-1000 HPC facility at the Physical Research Laboratory, Ahmedabad.

\end{acknowledgments}

\appendix

\section{\label{sec:appendixA}The $f(y)$ function}

In the conditions $|y|\ll1$ or $|y|\gg1$, the integral of the $f(y)$ function
\begin{equation}\label{eq:A1}
f(y) = \int_0^\infty \frac{y}{y^2 - x^2} \frac{x^3}{e^x - 1} \textrm{d}x
\end{equation}
can be simplified by using the definite integral formula
\begin{equation}\label{eq:A2}
\int_0^\infty \frac{x^{2n-1}}{e^{px} - 1} \textrm{d}x = (-1)^{n-1} \left(\frac{2\pi}{p}\right)^{2n} \frac{B_{2n}}{4n} ,
\end{equation}
where $B_{2n}$ is the Bernoulli number which can be expressed as
\begin{equation}\label{eq:A3}
B_n = \lim_{x\rightarrow0} \frac{\textrm{d}^n}{\textrm{d}x^n} \left(\frac{x}{e^x-1}\right) .
\end{equation}
For $|y|\ll1$, Eq. (\ref{eq:A1}) can be reduced to
\begin{eqnarray}
f(y) & \thickapprox &
\int_0^\infty -y \frac{x}{e^x - 1} \textrm{d}x  =  -y (\pi^2 B_2)  =  -\frac{\pi^2 y}{6}. \ \ \ \ \  \label{eq:A4}
\end{eqnarray}
For $|y|\gg1$, Eq. (\ref{eq:A1}) can be reduced to
\begin{eqnarray}
f(y) & \thickapprox &
\int_0^\infty \left( \frac{1}{y} + \frac{x^2}{y^3} + \cdots \right) \frac{x^3}{e^x - 1} \textrm{d}x \nonumber  \\
& = & \frac{1}{y} (-2\pi^4 B_4) + \frac{1}{y^3} (\frac{16\pi^6}{3} B_6) + \cdots \nonumber \\
& = & \frac{\pi^4}{15y} + \frac{8\pi^6}{63y^3} + \cdots.  \label{eq:A5}
\end{eqnarray}


\begin{thebibliography}{}

\bibitem{Bloom-Nature-2014}
B. J. Bloom, T. L. Nicholson, J. R. Williams, S. L. Campbell, M. Bishof, X. Zhang, W. Zhang, S. L. Bromley, and J. Ye, Nature \textbf{506}, 71 (2014).

\bibitem{Huntemann-PRL-2016}
N. Huntemann, C. Sanner, B. Lipphardt, Chr. Tamm, and E. Peik, Phys. Rev. Lett. \textbf{116}, 063001 (2016).

\bibitem{Schioppo-nphoton-2017}
M. Schioppo, R. C. Brown, W. F. McGrew, N. Hinkley, R. J. Fasano, K. Beloy, T. H. Yoon, G. Milani, D. Nicolodi, J. A. Sherman, N. B. Phillips, C. W. Oates, and A. D. Ludlow,
Nat. Photon. \textbf{11}, 48 (2017).

\bibitem{Brewer-PRL-2019}
S. M. Brewer, J.-S. Chen, A. M. Hankin, E. R. Clements, C. W. Chou, D. J. Wineland, D. B. Hume, and D. R. Leibrandt, Phys. Rev. Lett. \textbf{123}, 033201 (2019).

\bibitem{Boulder-Nature-2021}
Boulder Atomic Clock Optical Network (BACON) Collaboration, Nature \textbf{591}, 564 (2021).

\bibitem{Riehle-Metrologia-2018}
F. Riehle, P. Gill, F. Arias, and L. Robertsson,
Metrologia \textbf{55}, 188 (2018).

\bibitem{Milner-PRL-2019}
W. R. Milner, J. M. Robinson, C. J. Kennedy, T. Bothwell, D. Kedar, D. G. Matei, T. Legero, U. Sterr, F. Riehle,
H. Leopardi, T. M. Fortier, J. A. Sherman, J. Levine, J. Yao, J. Ye, and E. Oelker,
Phys. Rev. Lett. \textbf{123}, 173201 (2019).

\bibitem{McGrew-Nature-2018}
W. F. McGrew, X. Zhang, R. J. Fasano, S. A. Sch\"{a}ffer, K. Beloy, D. Nicolodi, R. C. Brown, N. Hinkley, G. Milani,
M. Schioppo, T. H. Yoon, and A. D. Ludlow,
Nature \textbf{564}, 87 (2018).

\bibitem{Schuldt-GPSsolut-2021}
T. Schuldt, M. Gohlke, M. Oswald, J. W\"{u}st, T. Blomberg, K. D\"{o}ringshoff, A. Bawamia, A. Wicht, M. Lezius,
K. Voss, M. Krutzik, S. Herrmann, E. Kovalchuk, A. Peters, and C. Braxmaier,
GPS Solut. \textbf{25} 83 (2021)

\bibitem{Burt-Nature-2021}
E. A. Burt, J. D. Prestage, R. L. Tjoelker, D. G. Enzer, D. Kuang, D. W. Murphy, D. E. Robison, J. M. Seubert,
R. T. Wang, and T. A. Ely,
Nature \textbf{595}, 43 (2021).

\bibitem{Kennedy-PRL-2020}
C. J. Kennedy, E. Oelker, J. M. Robinson, T. Bothwell, D. Kedar, W. R. Milner, G. E. Marti,
A. Derevianko, and J. Ye,
Phys. Rev. Lett. \textbf{125}, 201302 (2020).

\bibitem{Kobayashi-PRL-2022}
T. Kobayashi, A. Takamizawa, D. Akamatsu, A. Kawasaki, A. Nishiyama, K. Hosaka, Y. Hisai, M. Wada, H. Inaba,
T. Tanabe, and M. Yasuda,
Phys. Rev. Lett. \textbf{129}, 241301 (2022).

\bibitem{Lange-PRL-2021}
R. Lange, N. Huntemann, J. M. Rahm, C. Sanner, H. Shao, B. Lipphardt, Chr. Tamm, S. Weyers, and E. Peik,
Phys. Rev. Lett. \textbf{126}, 011102 (2021).

\bibitem{Safronova-RMP-2018}
M. S. Safronova, D. Budker, D. DeMille, D. F. Jackson Kimball, A. Derevianko, and C. W. Clark,
Rev. Mod. Phys. \textbf{90}, 025008 (2018).

\bibitem{Sanner-Nature-2019}
C. Sanner, N. Huntemann, R. Lange, C. Tamm, E. Peik, M. S. Safronova, and S. G. Porsev,
Nature \textbf{567}, 204 (2019).

\bibitem{Bothwell-Nature-2022}
T. Bothwell, C. J. Kennedy, A. Aeppli, D. Kedar, J. M. Robinson, E. Oelker, A. Staron, and J. Ye,
Nature \textbf{602}, 420 (2022).

\bibitem{Derevianko-RMP-2011}
A. Derevianko, and H. Katori,
Rev. Mod. Phys. \textbf{83}, 331 (2011).

\bibitem{Ludlow-RMP-2015}
A. D. Ludlow, M. M. Boyd, J. Ye, E. Peik, and P. O. Schmidt,
Rev. Mod. Phys. \textbf{87}, 637 (2015).

\bibitem{Kozlov-RMP-2018}
M. G. Kozlov, M. S. Safronova, J. R. Crespo L\'{o}pez-Urrutia, and P. O. Schmidt,
Rev. Mod. Phys. \textbf{90}, 045005 (2018).

\bibitem{Yu-review-2023}
Y.-M. Yu, B. K. Sahoo, and B.-B. Suo,
Front. Phys. \textbf{11}, 1104848 (2023).

\bibitem{King-Nature-2022}
S. A. King, L. J. Spie{\ss}, P. Micke, A. Wilzewski, T. Leopold, E. Benkler, R. Lange, N. Huntemann,
A. Surzhykov, V. A. Yerokhin, J. R. Crespo L\'{o}pez-Urrutia, and P. O. Schmidt,
Nature \textbf{611}, 43 (2022).

\bibitem{Derevianko-PRL-2012}
A. Derevianko, V. A. Dzuba, and V. V. Flambaum,
Phys. Rev. Lett. \textbf{109}, 180801 (2012).

\bibitem{Safronova-PRL-2014}
M. S. Safronova, V. A. Dzuba, V. V. Flambaum, U. I. Safronova, S. G. Porsev, and M. G. Kozlov,
Phys. Rev. Lett. \textbf{113}, 030801 (2014).

\bibitem{Yudin-PRL-2014}
V. I. Yudin, A. V. Taichenachev, and A. Derevianko,
Phys. Rev. Lett. \textbf{113}, 233003 (2014).

\bibitem{Yu-PRA-2016}
Y.-M. Yu, and B. K. Sahoo,
Phys. Rev. A \textbf{94}, 062502 (2016).

\bibitem{Yu-PRA-2018}
Y.-M. Yu, and B. K. Sahoo,
Phys. Rev. A \textbf{97}, 041403(R) (2018).

\bibitem{Yu-PRA-2019}
Y.-M. Yu, and B. K. Sahoo,
Phys. Rev. A \textbf{99}, 022513 (2019).

\bibitem{Bekker-ncommun-2019}
H. Bekker, A. Borschevsky, Z. Harman, C. H. Keitel, T. Pfeifer, P. O. Schmidt, J. R. Crespo L\'{o}pez-Urrutia,
and J. C. Berengut,
Nat. Commun. \textbf{10}, 5651 (2019).

\bibitem{Lyu-arxiv-2023}
C. Lyu, C. H. Keitel, and Z. Harman,
arXiv:2305.09603 (2023).

\bibitem{Yu-arxiv-2023}
Y.-M. Yu, and B. K. Sahoo,
arXiv:2307.14543 (2023).

\bibitem{Derevianko-QST-2022}
A. Derevianko, K. Gibble, L. Hollberg, N. R. Newbury, C. Oates, M. S. Safronova, L. C. Sinclair, and N. Yu,
Quantum Sci. Technol. \textbf{7}, 044002 (2022).

\bibitem{Rosenband-Science-2008}
T. Rosenband, D. B. Hume, P. O. Schmidt, C. W. Chou, A. Brusch, L. Lorini, W. H. Oskay, R. E. Drullinger, T. M. Fortier,
J. E. Stalnaker, S. A. Diddams, W. C. Swann, N. R. Newbury, W. M. Itano, D. J. Wineland, and J. C. Bergquist,
Science \textbf{319}, 1808 (2008).

\bibitem{Chou-PRL-2010}
C. W. Chou, D. B. Hume, J. C. J. Koelemeij, D. J. Wineland, and T. Rosenband,
Phys. Rev. Lett. \textbf{104}, 070802 (2010).

\bibitem{Rosenband-PRL-2007}
T. Rosenband, P. O. Schmidt, D. B. Hume, W. M. Itano, T. M. Fortier, J. E. Stalnaker, K. Kim,
S. A. Diddams, J. C. J. Koelemeij, J. C. Bergquist, and D. J. Wineland,
Phys. Rev. Lett. \textbf{98}, 220801 (2007).

\bibitem{Wei-CPB-2022}
Y.-F. Wei, Z.-M. Tang, C.-B. Li, Y. Yang, Y.-M. Zou, K.-F. Cui, and X.-R. Huang,
Chin. Phys. B \textbf{31}, 083102 (2022).

\bibitem{Kallay-PRA-2011}
M. K\'{a}llay, H. S. Nataraj, B. K. Sahoo, B. P. Das, and L. Visscher,
Phys. Rev. A \textbf{83}, 030503(R) (2011).

\bibitem{Safronova-PRL-2011}
M. S. Safronova, M. G. Kozlov, and C. W. Clark, Phys. Rev. Lett. \textbf{107}, 143006 (2011).

\bibitem{Ushijima-nphoton-2015}
I. Ushijima, M. Takamoto, M. Das, T. Ohkubo, and H. Katori,
Nat. Photon. \textbf{9}, 185 (2015).

\bibitem{Huang-PRAppl-2022}
Y. Huang, B. Zhang, M. Zeng, Y. Hao, Z. Ma, H. Zhang, H. Guan, Z. Chen, M. Wang, and K. Gao,
Phys. Rev. Appl. \textbf{17}, 034041 (2022).

\bibitem{Porsev-PRA-2006}
S. G. Porsev, and A. Derevianko,
Phys. Rev. A \textbf{74}, 020502(R) (2006).

\bibitem{Arora-PRA-2012}
B. Arora, D. K. Nandy, and B. K. Sahoo,
Phys. Rev. A. \textbf{85}, 012506 (2012).

\bibitem{Sahoo-PramanaJP-2014}
B. K. Sahoo, Pramana-J. Phys. \textbf{83}, 255 (2014).

\bibitem{Kajita-JPSJ-2019}
M. Kajita,
J. Phys. Soc. Jpn. \textbf{88}, 104001 (2019).

\bibitem{Farley-PRA-1981}
J. W. Farley, and W. H. Wing,
Phys. Rev. A \textbf{23}, 2397 (1981).

\bibitem{Dzuba-PRA-1996}
V. A. Dzuba, V. V. Flambaum, and M. G. Kozlov,
Phys. Rev. A \textbf{54}, 3948 (1996).

\bibitem{Kozlov-CPC-2015}
M. G. Kozlov, S. G. Porsev, M. S. Safronova, and I. I. Tupitsyn,
Comput. Phys. Comm. \textbf{195}, 199 (2015).

\bibitem{Fischer-JPB-2016}
C. Froese Fischer, M. Godefroid, T. Brage, P. J$\ddot{o}$nsson, and G. Gaigalas,
J. Phys. B: At. Mol. Opt. Phys. \textbf{49}, 182004 (2016).

\bibitem{Fischer-CPC-2019}
C. Froese Fischer, G. Gaigalas, P. J$\ddot{o}$nsson, and J. Biero$\acute{n}$,
Comput. Phys. Comm. \textbf{237}, 184 (2019).

\bibitem{Tang-JPB-2018}
Z.-M. Tang, Y.-M. Yu, J. Jiang, and C.-Z. Dong,
J. Phys. B: At. Mol. Opt. Phys. \textbf{51}, 125002 (2018).

\bibitem{Sturesson-CPC-2007}
L. Sturesson, P. J$\ddot{o}$nsson, and C. Froese Fischer,
Comput. Phys. Comm. \textbf{177}, 539 (2007).

\bibitem{NIST}
A. Kramida, Yu. Ralchenko, J. Reader, and NIST ASD Team (2022),
\textit{NIST Atomic Spectra Database} (ver. 5.10), https://physics.nist.gov/asd.

\bibitem{Martin-JPCRD-1979}
W. C. Martin, and R. Zalubas,
J. Phys. Chem. Ref. Data \textbf{8}, 817 (1979).

\bibitem{Brewer-PRA-2019}
S. M. Brewer, J.-S. Chen, K. Beloy, A. M. Hankin, E. R. Clements, C. W. Chou, W. F. McGrew, X. Zhang, R. J. Fasano,
D. Nicolodi, H. Leopardi, T. M. Fortier, S. A. Diddams, A. D. Ludlow, D. J. Wineland, D. R. Leibrandt, and D. B. Hume,
Phys. Rev. A \textbf{100}, 013409 (2019).

\bibitem{Guggemos-NJP-2019}
M. Guggemos, M. Guevara-Bertsch, D. Heinrich, O. A. Herrera-Sancho, Y. Colombe, R. Blatt, and C. F. Roos,
New J. Phys. \textbf{21}, 103003 (2019).

\bibitem{Hannig-RSI-2019}
S. Hannig, L. Pelzer, N. Scharnhorst, J. Kramer, M. Stepanova, Z. T. Xu, N. Spethmann, I. D. Leroux,
T. E. Mehlst\"{a}ubler, and P. O. Schmidt,
Rev. Sci. Instrum. \textbf{90}, 053204 (2019).

\bibitem{Ma-APB-2020}
Z. Y. Ma, H. L. Liu, W. Z. Wei, W. H Yuan, P. Hao, Z. Deng, H. Che, Z. T. Xu, F. H. Cheng, Z. Y. Wang, K. Deng,
J. Zhang, and Z. H. Lu,
Appl. Phys. B \textbf{126}, 129 (2020).

\bibitem{Shao-CPL-2019}
S.-J. Chao, K.-F. Cui, S.-M. Wang, J. Cao, H.-L. Shu, X.-R. Huang,
Chin. Phys. Lett. \textbf{36}, 120601 (2019).

\bibitem{Cui-EPJD-2022}
K. Cui, S. Chao, C. Sun, S. Wang, P. Zhang, Y. Wei, J. Yuan, J. Cao, H. Shu, and X. Huang,
Eur. Phys. J. D \textbf{107}, 143006 (2022).

\bibitem{Middelmann-PRL-2012}
T. Middelmann, S. Falke, C. Lisdat, and U. Sterr,
Phys. Rev. Lett. \textbf{109}, 263004 (2012).

\bibitem{Wu-arxiv-2023-1}
F.-F. Wu, T.-Y. Shi, and L.-Y. Tang,
arXiv:2301.06740 (2023).

\bibitem{Wu-arxiv-2023-2}
F.-F. Wu, T.-Y. Shi, W.-T. Ni, and L.-Y. Tang,
arXiv:2306.08414 (2023).

\bibitem{Porsev-arxiv-2023}
S. G. Porsev, M. G. Kozlov, and M. S. Safronova,
arXiv:2306.10173 (2023).

\bibitem{Wu-NJP-2023}
L. Wu, X. Wang, T. Wang, J. Jiang, and C. Dong,
New J. Phys. \textbf{25}, 043011 (2023).

\bibitem{Kulosa-PRL-2015}
A. P. Kulosa, D. Fim, K. H. Zipfel, S. R\"{u}hmann, S. Sauer, N. Jha, K. Gibble, W. Ertmer, E. M. Rasel,
M. S. Safronova, U. I. Safronova, and S. G. Porsev,
Phys. Rev. Lett. \textbf{115}, 240801 (2015).

\bibitem{Takamoto-Nature-2005}
M. Takamoto, F.-L. Hong, R. Higashi, and H. Katori,
Nature \textbf{435}, 321 (2005).

\bibitem{Yamaguchi-PRL-2019}
A. Yamaguchi, M. S. Safronova, K. Gibble, and H. Katori,
Phys. Rev. Lett. \textbf{123}, 113201 (2019).

\bibitem{Ohtsubo-Hyperfine-2019}
N. Ohtsubo, Y. Li, K. Matsubara, N. Nemitz, H. Hachisu, T. Ido, and K. Hayasaka,
Hyperfine Interact. \textbf{240}, 39 (2019).

\bibitem{Lemke-PRL-2009}
N. D. Lemke, A. D. Ludlow, Z. W. Barber, T. M. Fortier, S. A. Diddams, Y. Jiang, S. R. Jefferts, T. P. Heavner,
T. E. Parker, and C. W. Oates,
Phys. Rev. Lett. \textbf{103}, 063001 (2009).

\bibitem{McFerran-PRL-2012}
J. J. McFerran, L. Yi, S. Mejri, S. Di Manno, W. Zhang, J. Gu\'{e}na, Y. Le Coq, and S. Bize,
Phys. Rev. Lett. \textbf{108}, 183004 (2012).

\bibitem{Chwalla-PRL-2009}
M. Chwalla, J. Benhelm, K. Kim, G. Kirchmair, T. Monz, M. Riebe, P. Schindler, A. S. Villar, W. H\"{a}nsel,
C. F. Roos, R. Blatt, M. Abgrall, G. Santarelli, G. D. Rovera, and Ph. Laurent,
Phys. Rev. Lett. \textbf{102}, 023002 (2009).

\bibitem{Huang-PRL-2016}
Y. Huang, H. Guan, P. Liu, W. Bian, L. Ma, K. Liang, T. Li, and K. Gao,
Phys. Rev. Lett. \textbf{116}, 013001 (2016).

\bibitem{Margolis-Science-2004}
H. S. Margolis, G. P. Barwood, G. Huang, H. A. Klein, S. N. Lea, K. Szymaniec, and P. Gill,
Science \textbf{306}, 1355 (2004).

\bibitem{Steinel-PRL-2023}
M. Steinel, H. Shao, M. Filzinger, B. Lipphardt, M. Brinkmann, A. Didier, T. E. Mehlst\"{a}ubler,
T. Lindvall, E. Peik, and N. Huntemann,
Phys. Rev. Lett. \textbf{131}, 083002 (2023).

\bibitem{Arnold-PRL-2020}
K. J. Arnold, R. Kaewuam, S. R. Chanu, T. R. Tan, Z. Zhang, and M. D. Barrett,
Phys. Rev. Lett. \textbf{124}, 193001 (2020).

\bibitem{Holliman-PRL-2022}
C. A. Holliman, M. Fan, A. Contractor, S. M. Brewer, and A. M. Jayich,
Phys. Rev. Lett. \textbf{128}, 033202 (2022).

\bibitem{Roberts-PRA-1999}
M. Roberts, P. Taylor, S. V. Gateva-Kostova, R. B. M. Clarke, W. R. C. Rowley, and P. Gill,
Phys. Rev. A \textbf{60}, 2867 (1999).

\bibitem{Oskay-PRL-2006}
W. H. Oskay, S. A. Diddams, E. A. Donley, T. M. Fortier, T. P. Heavner, L. Hollberg, W. M. Itano, S. R. Jefferts,
M. J. Delaney, K. Kim, F. Levi, T. E. Parker, and J. C. Bergquist,
Phys. Rev. Lett. \textbf{97}, 020801 (2006).

\bibitem{Fan-PRL-2023}
X. Fan, T. G. Myers, B. A. D. Sukra, and G. Gabrielse,
Phys. Rev. Lett. \textbf{130}, 071801 (2023).

\bibitem{Safronova-PRL-2018}
M. S. Safronova, S. G. Porsev, C. Sanner, and J. Ye,
Phys. Rev. Lett. \textbf{120}, 173001 (2018).

\bibitem{Yamaguchi-NJP-2010}
A. Yamaguchi, S. Uetake, S. Kato, H. Ito, and Y. Takahashi,
New J. Phys. \textbf{12}, 103001 (2010).

\bibitem{Dzuba-PRA-2018-yb}
V. A. Dzuba, V. V. Flambaum, and S. Schiller,
Phys. Rev. A \textbf{98}, 022501 (2018).

\bibitem{Ishiyama-PRL-2023}
T. Ishiyama, K. Ono, T. Takano, A. Sunaga, and Y. Takahashi,
Phys. Rev. Lett. \textbf{130}, 153402 (2023).

\bibitem{Kawasaki-PRA-2023}
A. Kawasaki, T. Kobayashi, A. Nishiyama, T. Tanabe, and M. Yasuda,
Phys. Rev. A \textbf{107}, L060801 (2023).

\bibitem{Tang-PRA-2023}
Z.-M. Tang, Y.-M Yu, B. K. Sahoo, C.-Z. Dong, Y. Yang, and Y. Zou,
Phys. Rev. A \textbf{107}, 053111 (2023).

\bibitem{Schneider-PRL-2005}
T. Schneider, E. Peik, and Chr. Tamm,
Phys. Rev. Lett. \textbf{94}, 230801 (2005).

\bibitem{Huntemann-PRL-2012}
N. Huntemann, M. Okhapkin, B. Lipphardt, S. Weyers, Chr. Tamm, and E. Peik,
Phys. Rev. Lett. \textbf{108}, 090801 (2012).

\bibitem{Furst-PRL-2020}
H. A. F\"{u}rst, C.-H. Yeh, D. Kalincev, A. P. Kulosa, L. S. Dreissen, R. Lange, E. Benkler,
N. Huntemann, E. Peik, and T. E. Mehlst\"{a}ubler,
Phys. Rev. Lett. \textbf{125}, 163001 (2020).

\bibitem{Arnold-ncommun-2018}
K. J. Arnold, R. Kaewuam, A. Roy, T. R. Tan, and M. D. Barrett,
Nat. Commun. \textbf{9}, 1650 (2018).

\bibitem{Zhiqiang-sciadv-2023}
Z. Zhiqiang, K. J. Arnold, R. Kaewuam, M. D. Barrett,
Sci. Adv. \textbf{9}, eadg1971 (2023).

\end{thebibliography}
\end{document}